\documentclass[times]{qjrms4}

\usepackage[T1]{fontenc}
\usepackage[utf8]{inputenc}
\usepackage{textcomp}   % degree sign used in the domain figure caption
\usepackage{xspace}
\usepackage{physics}
\usepackage{mathtools}
\usepackage{bm}
\usepackage{url}
\usepackage[mathlines]{lineno}

\newcommand{\ie}{\textit{i}.\textit{e}.,\@\xspace}
\newcommand{\eg}{\textit{e}.\textit{g}.,\@\xspace}
\newcommand{\vs}{\textit{vs}.\@\xspace}
\newcommand{\eq}{Eq.\@\xspace}
\newcommand{\eqs}{Eqs.\@\xspace}
\newcommand{\fig}{Fig.\@\xspace}
\newcommand{\figs}{Figs.\@\xspace}

\newcommand*\mean[1]{\mkern 1.5mu\overline{\mkern-1.5mu#1\mkern-1.5mu}\mkern 1.5mu}
\newcommand{\mat}[1]{\mathbf{#1}}
\newcommand{\bv}[1]{\bm{#1}}
\newcommand{\oper}[1]{\mathcal{#1}}
\newcommand{\newset}[1]{\mathbb{#1}}
\providecommand{\diag}{\operatorname{diag}}

\providecommand{\doi}[1]{doi:\discretionary{}{}{}\url{#1}}

\def\volumeyear{2026}

\usepackage{etoolbox}
\makeatletter
\patchcmd{\@maketitle}
  {\mbox{}\hfill\framebox[43mm]{\rule{0pt}{15mm}}\hfill\mbox{}}
  {}
  {\typeout{QJ-PATCH: logo placeholder box removed}}
  {\PackageWarning{PEF}{QJ-PATCH FAILED: logo placeholder box still present}}
\makeatother

\makeatletter
\g@addto@macro\ps@qjrms{%
  \def\@evenfoot{\parbox[t]{\textwidth}{\footnotesize\noindent
    \copyrightline\hfill\emph{Prepared using \textsf{qjrms4.cls}}}}%
  \let\@oddfoot\@evenfoot}
\g@addto@macro\ps@title{%
  \def\@oddfoot{\parbox[t]{\textwidth}{\footnotesize\noindent
    \copyrightline\hfill\textit{Prepared using \textsf{qjrms4.cls} [Version: \update]}}}%
  \let\@evenfoot\@oddfoot}
\makeatother
\begin{document}
% \linenumbers

\runningheads{J.~P.~Brook \etal}{Preemptive forecasting of radar reflectivity}

\title{Rapid Corrections of Ensemble Forecasts using Radar Imagery ---\\
Preemptive Forecasting Using Flow-dependent Covariance Localization}

\author{Jordan P. Brook\affil{a,b}\corrauth, Craig H. Bishop\affil{a,c,d} and
Alain Protat\affil{a}}

\address{\affilnum{a}Science and Innovation Group, Bureau of Meteorology, Australia\\
\affilnum{b}School of the Environment, The University of Queensland, Australia\\
\affilnum{d}School of Geography, Earth and Atmospheric Science, University of Melbourne, Australia\\
\affilnum{d}ARC Centre of Excellence for 21st Century Weather}

\corraddr{J. Brook, School of the Environment, The University of Queensland, Australia.
E-mail: j.brook@uq.edu.au}

\begin{abstract}
%%% ABSTRACT COPIED VERBATIM FROM THE AMS MANUSCRIPT.
%%% QJRMS limit is 300 words and it must contain no citations -- see TODO_QJRMS.md.
The full data-assimilation forecast cycle is computationally expensive. Consequently, observations used to initialize forecasts are often several hours old by the time they are assimilated. To best assess the current state of the atmosphere, forecasters may identify which members of the latest ensemble forecast most accurately track real-time observations and prefer those members accordingly. In this study, we formalize this ensemble weighting procedure using the Ensemble Transform Kalman Filter (ETKF) with flow dependent ensemble covariance localization to create a “preemptive” forecast – an updated forecast generated without rerunning the expensive numerical weather prediction (NWP) model. We present this novel methodology and evaluate it in a challenging and operationally significant application: precipitation nowcasting with weather radar fields. Validation is conducted using data from the Australian radar network and the Bureau of Meteorology's operational convective-scale limited-area model (ACCESS-CE). Both deterministic and ensemble forecasts are evaluated and show statistically significant improvements over the background NWP for the first 2--6 hours of probabilistic precipitation forecasts over a 9-week trial period.
\end{abstract}

%%% TODO: QJRMS requires 3--8 keywords, at least two of which appear in the
%%% title, separated by semicolons.  Draft set below -- please review.

\maketitle

\section{Introduction}\label{intro}

Numerical weather forecasting is a computationally intensive task that has evolved alongside expanding computational resources over the past century. Short-term nowcasting (0--2 h) of highly unpredictable severe weather events has benefited enormously from the implementation of rapidly updating models that directly resolve convection, such as the ICON-Rapid Update Cycle \citep[ICON-RUC;][]{Zangl15} and the Warn On Forecast System \citep[WOFS;][]{Stensrud13}. These systems harness immense computing power to run a high-resolution ensemble forecast with data assimilation (DA) updates on sub-hourly scales. In this way, such systems can represent and forecast the most rapidly evolving atmospheric phenomena such as severe convective thunderstorms. 

In operational weather forecasting centers with large domains and limited computational resources, such as Australia, running a large rapidly updating convection-permitting ensemble using physics-based models remains unfeasible. Instead, convective-scale ensemble forecasts are run with 12 ensemble members every 6 hours, with hourly DA updates \citep{Rennie22}. Both the data assimilation and forecast update cycles are too infrequent to faithfully represent a highly unstable atmosphere as it evolves in real time. As a result, forecasts mostly provide qualitative guidance---\ie is there a chance of supercells today?---rather than quantifying the exact timing, location, or intensity of such events. In this study, we attempt to improve the quantitative behavior of these short-term forecasts by implementing \citeauthor{Etherton07}'s \citeyearpar{Etherton07} preemptive forecasting technique and overcoming some of its limitations by developing new variants of \citeauthor{Bishop07}'s \citeyearpar{Bishop07, Bishop09} flow-adaptive ensemble covariance localization techniques. The new approach expresses the localized ensemble covariances in terms of a flow evolving modulated ensemble \citep{Bishop09} that is far larger than the original ensemble. This increase in ensemble size enables information from a larger number of observations  to be captured and projected forward in time than \citeauthor{Etherton07}'s \citeyearpar{Etherton07} suggested approach.

Preemptive Forecasting works by propagating data assimilation updates from the analysis time to later lead times using information from an existing ensemble forecast. This type of 4D\footnote{Here, we use the colloquial definition of the term `4D', meaning a state with both spatial and temporal dimensions, rather than a state containing specifically four dimensions.} assimilation technique adjusts future forecast states without rerunning the nonlinear weather model. In practice, this capability is most useful when a fast, observationally informed adjustment is needed between expensive model cycles. Section \ref{pef} outlines how preemptive forecasts are only identical to actual forecasts given an infinitely large ensemble governed by linear dynamics. However, previous studies have also shown its utility in practical scenarios such as forecasting large-scale phenomena with a 64-member ensemble \citep{Etherton07}. Here, we have chosen the more difficult, operationally relevant task of assimilating radar reflectivity imagery into a 12-member ensemble\footnote{In practice we create a 24-member ``lagged'' ensemble from the two most recent 12-member forecasts.} forecast to improve short-term predictions of precipitation. This extends preemptive forecasting to smaller ensemble sizes and more highly nonlinear dynamics than those guaranteed by theory or previously attempted in the literature. 

Further to the theoretical challenges of preemptive forecasting, our task also inherits all the known challenges of standard data assimilation of radar imagery. These challenges arise when hydrometeors are initially poorly forecast, and prove difficult to correct with radar data due to a combination of factors, including: observations being nonlinearly related to prognostic model variables, exhibiting strong small-scale structures, evolving rapidly over time, and being characterized by highly nonstationary error covariances. Two practical problems arise due to these characteristics: 1) high-frequency errors due to sampling biases in small ensembles can dominate DA increments \citep{Berre10}, and 2) precipitation is frequently observed at grid points where none is forecast in any ensemble member (\ie zero ensemble spread). In this study, we introduce a new variant of \citeauthor{Bishop07}'s \citeyearpar{Bishop07, Bishop09} localization methodology to alleviate these issues and enable skillful preemptive forecasts. We expect these methodological contributions will also be useful for preemptively forecasting dense imagery of any kind (\eg radar, satellite).

The aim of this manuscript is to introduce the above methodology and briefly demonstrate it successfully improves probabilistic forecasts of precipitation in operational model fields. We demonstrate this using historically archived cases from the convection-permitting, operational, limited-area, ensemble version of the Australian Community Climate and Earth-System Simulator (ACCESS) model, named ACCESS-CE \citep{Rennie22, Cooper25}. The remainder of this manuscript is structured as follows: Section~\ref{method} introduces ensemble data assimilation, preemptive forecasting, and other methodological contributions; Section~\ref{results} demonstrates the application of these contributions to an operational forecast model; Section~\ref{results2} quantifies the performance of the technique on a large dataset of archival cases; and Section~\ref{conclusion} summarizes the study and outlines future work. 

\section{Method}\label{method}

\subsection{Notation and Theory}\label{etkf}

In this study, we follow the standard notation of \citet{Ide97} by denoting an ensemble of $k$ model forecasts as follows:

\begin{equation}
    \mat{X}^f = \left[\vb{x}^f_1, \vb{x}^f_2, \dots, \vb{x}^f_k \right] \in \newset{R}^{n\times k},
\end{equation}
\noindent
 where $n$ is the total size of each forecast state vector. Note that bold font upper- and lower-case variables represent matrices and vectors, respectively. In the notation that follows, it is also convenient to express the ensemble forecast as the sum of its mean and perturbations: 

\begin{equation}\label{pert}
    \mat{X}^f = \mean{\mat{X}^f} + \left(\sqrt{k-1}\right)\mat{Z}^f,
\end{equation}
\noindent
where $\mean{\mat{X}^f}$ consists of $k$ repeated columns of $\mean{\vb{x}^f}$, and 

\noindent
\begin{equation}\label{Z}
\mat{Z}^f=\left(\vb{x}^f_1 - \mean{\vb{x}^f}, \dots, \vb{x}^f_k - \mean{\vb{x}^f}\right)\left. \middle/ \sqrt{k-1} \right.\in \newset{R}^{n\times k}.
\end{equation} 
\noindent 
The ensemble perturbations defined above importantly represent the square root of the sample forecast error covariance matrix:

\begin{equation}\label{cov}
\mat{P}^f = \mat{Z}^f (\mat{Z}^f)^T \in \newset{R}^{n\times n}.
\end{equation}

Ensemble Kalman filters (EnKFs) use this sample matrix in \eq \ref{cov} to estimate covariances between forecast errors at different locations, model variables, and, in our case, time steps. This enables the forecast to be combined with a set of $p$ observations, denoted $\vb{y} \in \newset{R}^p$, to produce an analysis ensemble $\mat{X}^a$. There exist a number of EnKF methodologies in the literature, each calculating $\mat{X}^a$ in a different manner. The earliest methods were stochastic EnKFs which produced an analysis ensemble by assimilating perturbed observations \citep{Houtekamer98}, followed later by deterministic EnKFs such as the Ensemble Transform Kalman Filter \citep[ETKF;][]{Bishop01}, the Ensemble Adjustment Kalman Filter \citep[EAKF;][]{Anderson01}, and the Ensemble Square Root Filter \citep[EnSRF;][]{Whitaker02}. The ETKF was selected in this study. 

The ETKF is preferred here for a number of reasons, including its computational efficiency \citep{Tippett03}, its use in operational weather forecasting settings \citep{Schraff16, Whitaker22}, and its amenability to the preemptive forecasting approach, outlined in Section~\ref{pef}. The analysis is derived through a transformation into a reduced-rank subspace, which \citet{Posselt12} showed may be taken as the smallest dimension of either the ensemble size ($k$) or the number of observations assimilated ($p$). The decomposition for each case is formalized below:
\begin{subequations}\label{svd} 
\begin{gather} 
\makebox[\displaywidth][l]{\ensuremath{\text{if } k \le p:}}\notag\\ 
(\mat{H}\mat{Z})^T \mat{R}^{-1} \mat{H}\mat{Z} = \mat{C}_k\mat{\Gamma}\mat{C}_k^T, \label{eq:branches-a}\\ 
\makebox[\displaywidth][l]{\ensuremath{\text{else if } k > p:}}\notag\\ 
\mat{R}^{-1/2}\mat{H}\mat{Z}(\mat{R}^{-1/2}\mat{H}\mat{Z})^T = \mat{E}\mat{\Lambda}\mat{E}^T, \label{eq:branches-b}\\
\mat{C}_p = (\mat{R}^{-1/2}\mat{H}\mat{Z})^T\mat{E}\mat{\Lambda}^{-1/2}, \label{eq:branches-c}
\end{gather} 
\end{subequations} 

\noindent where $\mat{\Gamma}, \mat{C}_k \in \newset{R}^{k\times k}$ and $\mat{\Lambda}, \mat{E} \in \newset{R}^{p\times p}$ are the singular value and singular vector matrices, respectively, for each case, and $\mat{C}_p \in \newset{R}^{k\times p}$.

Having selected the appropriate SVD from \eq \eqref{svd}, the forecast mean and perturbations are updated without explicitly computing the $n\times n$ forecast error covariance matrix as follows:

\begin{equation}\label{etkf_mean}
\mean{\bv{x}^a}
=
\begin{cases}
\mean{\bv{x}^f} + \mat{Z}^f\mat{C}_k(\mat{\Gamma}+\mat{I} )^{-1}\mat{C}_k^T (\mat{H}\mat{Z}^f)^T\mat{R}^{-1} (\bv{y}-\mean{\mat{H}\bv{x}^f}), & \hspace{-.1cm} \text{if } k \le p, \\
\mean{\bv{x}^f} + \mat{Z}^f\mat{C}_p(\mat{\Lambda}+\mat{I} )^{-1}\mat{C}_p^T (\mat{H}\mat{Z}^f)^T\mat{R}^{-1} (\bv{y}-\mean{\mat{H}\bv{x}^f}), & \hspace{-.1cm} \text{if } k > p
\end{cases}
\end{equation}

\begin{equation}\label{etkf_pert}
\mat{Z}^a
=
\begin{cases}
\mat{Z}^f\mat{C}_k(\mat{\Gamma}+\mat{I} )^{-1/2}\mat{C}_k^T, & \hspace{-.1cm} \text{if } k \le p, \\
\mat{Z}^f-\mat{Z}^f\mat{C}_p\left[\mat{I} - (\mat{\Lambda}+\mat{I} )^{-1/2}\right]\mat{C}_p^T, & \hspace{-.1cm} \text{if } k > p
\end{cases}
\end{equation}

Once the analysis mean and perturbations have been calculated using \eqs \eqref{etkf_mean} \& \eqref{etkf_pert}, the analysis ensemble may then be constructed in the manner of \eq \eqref{pert}. The interested reader is referred to several comprehensive review articles for a complete derivation of the ETKF grounded in Bayesian theory \citep[\eg][]{Vetra-Carvalho18}.

\subsection{Preemptive Forecasting}\label{pef}

In this section we describe how the standard ETKF from Section~\ref{etkf} may also be used to produce an analysis not just at the time the observations are taken, but for all future times in the ensemble. This type of observationally-informed forecast adjustment is termed a ``preemptive'' forecast, \ie a forecast produced without rerunning the nonlinear model \citep{Etherton07}. Preemptive forecasting necessarily involves a `4D' forecast state: $\bv{x}^f \in \newset{R}^n$, where $n = n_i \times n_t$ becomes the product of the number of model variables at the initial time\footnote{Despite the limited focus on 2D, single variable forecasts of radar reflectivity in this study, the proposed methodology is generalizable to any number of model variables and spatial dimensions.} ($n_i$) and the number of time steps in the forecast ($n_t$). 

\eq \eqref{etkf_mean} gives the analysis mean in terms of the forecast at the initial time, which may be rewritten as:

\begin{equation}
    \mean{\bv{x}^a_{(i)}} = \mean{\bv{x}^f_{(i)}} + \mat{Z}^f_{(i)}\bv{w},
\end{equation}
\noindent 
where $\vb{w} = \mat{C}(\mat{\Gamma} + \mat{I})^{-1}\mat{C}^T (\mat{H}\mat{Z}^f_{(i)})^T\mat{R}^{-1} (\bv{y}-\mean{\mat{H}\bv{x}^f})$ and subscripts within parentheses denote time. 

To propagate the analysis state forward to a future time $t = F$ without rerunning the model $\oper{M}$, we make the approximation:

\begin{equation}\label{assumption}
    \oper{M}_{(i:F)}\left(\mean{\bv{x}^a_{(i)}}\right) \approx \mean{\bv{x}^f_{(F)}} + \mat{Z}^f_{(F)}\bv{w},
\end{equation}
\noindent
where the subscript notation for $\oper{M}$ indicates the model operates from time $i$ to $F$. 
This assumption is exact when the ensemble perturbations are governed by linear dynamics, because in that case 

% \begin{align}\label{deriv}
%     \oper{M}_{(i:F)}\left(\mean{\bv{x}^a_{(i)}}\right) &= \oper{M}_{(i:F)}\left(\mean{\bv{x}^f_{(i)}} + \mat{Z}^f_{(i)}\bv{w}\right)\nonumber\\
%     &= \oper{M}_{(i:F)}\left(\mean{\bv{x}^f_{(i)}}\right) + \left[\oper{M}_{(i:F)}\left(\mean{\bv{x}^f_{(i)}} + \mat{Z}^f_{(i)}\bv{w}\right) - \oper{M}_{(i:F)}\left(\mean{\bv{x}^f_{(i)}}\right)\right]\nonumber\\
%     &\approx \oper{M}_{(i:F)}\left(\mean{\bv{x}^f_{(i)}}\right) + \mat{M}_{(i:F)}\mat{Z}^f_{(i)}\bv{w} \nonumber\\
%     &\approx \mean{\bv{x}^f_{(F)}} + \mat{Z}^f_{(F)}\bv{w},
% \end{align}
\begin{alignat}{2}\label{deriv}
    \oper{M}_{(i:F)}\left(\mean{\bv{x}^a_{(i)}}\right)
    &= \oper{M}_{(i:F)}\left(
        \mean{\bv{x}^f_{(i)}} + \mat{Z}^f_{(i)}\bv{w}
    \right) && \nonumber\\
    &= \oper{M}_{(i:F)}\left(\mean{\bv{x}^f_{(i)}}\right)
       + && \nonumber\\
    && \mathllap{
        \left[
        \oper{M}_{(i:F)}\left(
            \mean{\bv{x}^f_{(i)}} + \mat{Z}^f_{(i)}\bv{w}
        \right)
        - \oper{M}_{(i:F)}\left(\mean{\bv{x}^f_{(i)}}\right)
        \right]
    } & \nonumber\\
    &\approx \oper{M}_{(i:F)}\left(\mean{\bv{x}^f_{(i)}}\right)
        + \mat{M}_{(i:F)}\mat{Z}^f_{(i)}\bv{w}
        && \nonumber\\
    &\approx \mean{\bv{x}^f_{(F)}} + \mat{Z}^f_{(F)}\bv{w},
        &&
\end{alignat}

\noindent 
where, by definition, the tangent linear model $\mat{M}_{(i:F)}$ satisfies

\begin{equation}\label{tangent_linear}
    \left[\oper{M}_{(i:F)}\left(\mean{\bv{x}^f_{(i)}} + \mat{Z}^f_{(i)}\bv{w}\right)- \oper{M}_{(i:F)}\left(\mean{\bv{x}^f_{(i)}}\right)\right] = \mat{M}_{(i:F)}\mat{Z}^f_{(i)}\bv{w},
\end{equation}
\noindent
when $\mat{Z}^f_{(i)}\bv{w}$ is infinitesimally small. Similar logic may also be used to express the ensemble perturbations at time $F$ by approximating \eq \eqref{etkf_pert} as follows:

\begin{equation}
    \mat{Z}^a_{(F)} = \mat{Z}^f_{(F)}\mat{C}\left(\mat{I}+\mat{\Gamma}\right)^{-1/2}\mat{C}^T.
\end{equation}

Once again, the analysis mean and perturbations at time $F$ may be combined using equation \eqref{pert} to produce the full ensemble preemptive forecast. 

It is important to note that the linear assumption made in \eq \eqref{assumption} may not hold in many practical applications. As a result, the updates made to the forecast may actually degrade its initial accuracy. In other words, the preemptive forecast only provably improves the initial forecast in the limit of linear dynamics. However, the aim of this study is to show that despite these theoretical shortcomings, preemptive forecasting can be a useful tool in practice, even for highly non-linear weather regimes and observation operators. 

\subsection{Flow-Adaptive Localization}\label{sec:localization}

\begin{figure*}[ht]
    \centering
    \noindent\includegraphics[width=\textwidth]{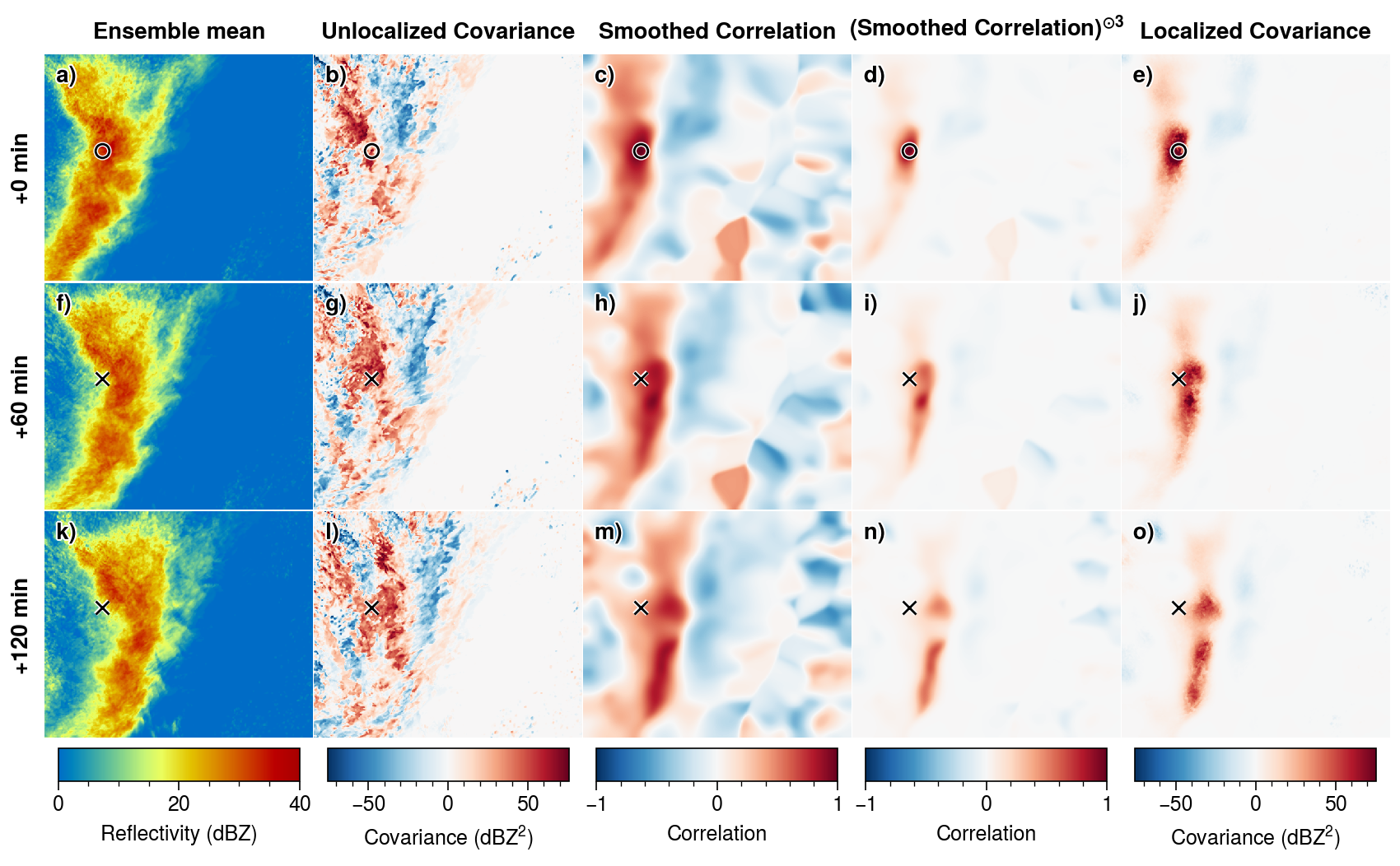}\\
    \caption{An illustration of the localization procedure on 0100 UTC 8 Sep 2022 for a single grid point ($i$, black circle at +0 min). This location is also marked at $+60$, and $+120$ min with a cross. The ensemble-mean reflectivity is shown for context in column 1, alongside row $i$ of the unlocalized 24-member forecast error covariance $\mat{P}^f_i$, in column 2. Column 2 thereby shows how \textit{i}, the single reflectivity grid point at time $t=0$ marked by the black circle, is correlated to others in space and time. The smoothed correlation $\mat{\Omega}^f$ is shown similarly in column 3, along with its elementwise cube $(\mat{\Omega}^f)^{\odot 3}$ in column 4, and the resulting localized covariance $\mat{P}_{\mathrm{loc}}^f$ in column 5. Cubing the smoothed correlation suppresses the spurious long-range structure in the raw covariance, yielding a compact, flow-dependent localized covariance.}\label{loc}
\end{figure*}

The accuracy of ensemble Kalman filters is largely contingent on the extent to which a finite-size ensemble can approximate the unknown forecast error covariance matrix. This is particularly difficult for high-resolution radar assimilation, where a very large ensemble size is theoretically required to constrain the small-scale variability present in the observations. Figure \ref{loc} has been included to demonstrate an example of the localization procedure, and illustrates the presence of significant long-range, texturally spurious correlations in the unlocalized covariance matrix (column 2, \figs \ref{loc}b,\ref{loc}g, and \ref{loc}l). These spurious correlations are not physically meaningful, and, as a result, preemptive forecasting experiments with this ensemble yielded extremely poor analyses (not shown). Standard practice dictates that localization must be applied to mitigate their harmful influence. 

It is important to note that common distance-based localization techniques \citep[\eg LETKF;][]{Hunt07} are not immediately suited to preemptive forecasting due to the added temporal dimension. Here, it is unsafe to assume that current observations covary strongest with proximal points at future times due to the movement of weather systems. One could attempt to re-center the future localization radius by tracking this movement (\ie advective localization); however, we estimate such an approach is too inefficient and error-prone for experiments such as these on large model fields. Rather, we turn to a flow-adaptive method that naturally localizes covariances through time, namely, a novel variation of the SENCORP technique \citep[Smoothed ENsemble COrrelations Raised to a Power;][]{Bishop07}. 

Flow-adaptive localization via SENCORP requires the application of a spatial smoother. Here, we employ a Gaussian low-pass filter in spectral space using the kernel $S(k)=\exp\!\big(\!-k^{2}/2\ell^{2}\big)$ for radial wavenumber $k$ and tunable smoothing scale $\ell$. The spectral transformation is performed via the discrete cosine transform (DCT), whose implicit mirror-symmetric boundary extension does not introduce spurious edge effects in a non-periodic limited-area domain \citep{Denis02}. 

Following \citet{Bishop07}, the symbol \(\odot\) denotes the Schur, or elementwise, product of two matrices. This notation is also extended to elementwise powers, such that
\begin{equation}
    \mat{A}^{\! \odot x} = \underset{1}{\mat{A}}\odot\underset{2}{\mat{A}}\dots\odot\underset{x}{\mat{A}}.
\end{equation}

Let \(\mat{Z}^f\in\newset{R}^{n\times k}\) denote the normalized forecast perturbation matrix defined in \eqref{Z}, and let \(\mat{Z}^f_s\in\newset{R}^{n\times k}\) denote the corresponding perturbation matrix after applying the DCT smoother. Now, the unsmoothed and smoothed ensemble variances are defined as

\begin{equation}
    \bv{\sigma}
    =
    \left(\mat{Z}^f\right)^{\! \odot 2}\bv{1}_k,
    \qquad
    \bv{\omega}
    =
    \left(\mat{Z}^f_s\right)^{\! \odot 2}\bv{1}_k,
    \label{loc_sigma}
\end{equation}
where a length-\(k\) vector of ones, \(\bv{1}_k\), is shorthand for a summation taken over the ensemble (column) dimension so that $\bv{\sigma}, \bv{\omega} \in\newset{R}^{n\times 1}$.

Initially, the smoothed variances \(\bv{\omega}\) are used to define the normalized, smoothed ensemble perturbations $\mat{W}^f$
\begin{equation}
    \mat{W}^f = \diag\left((\bv{\omega}+\epsilon)^{-1/2}\right)\mat{Z}^f_s%\big/\diag\left(\sqrt{\bv{\omega} + \epsilon}\right),
    \label{loc_W}
\end{equation}
\noindent 
where $\epsilon$ is a very small positive constant to avoid division by zero. Rows corresponding to smoothed variances $\omega\gg \epsilon$ have approximately unit norm, and represent the square root of the smoothed, flow-dependent forecast error correlation matrix,
\begin{equation}\label{corr}
    \mat{\Omega}^f = \mat{W^f}(\mat{W^f})^T.
\end{equation}

The smoothed correlation matrix $\mat{\Omega}^f$ varies from approximately unity along its diagonal\footnote{Note that \figs \ref{loc}c, \ref{loc}h, and \ref{loc}m show a single row of the correlation matrix and the circled point sits on its diagonal.} when $\omega \gg \epsilon$ to anywhere between -1 and 1 on off-diagonals (see \figs \ref{loc}c, \ref{loc}h, and \ref{loc}m). Note that a significant portion of spurious, small-scale correlations have been removed by the smoothing, revealing large-scale structures obscured by sampling noise in the unlocalized covariance (\figs \ref{loc}b, \ref{loc}g, and \ref{loc}l). The idea behind SENCORP-like localization techniques is to minimize weak correlations $\left(\abs{\mat{\Omega}^f}\ll1\right)$ by raising the correlation matrix to an elementwise power. Raising to a power leaves strong correlations $\approx 1$ roughly unchanged, thereby retaining only reliable, short-range correlations. Note that the elementwise power operation is unsuitable on its own as it significantly narrows the ensemble's correlation length scale. Therefore, smoothing (broadening) is performed before the elementwise power, removing weak correlations while preserving the correlation length scale of the original ensemble. 

We now employ the ensemble modulation product to implement this elementwise power through the ensemble perturbations, thereby avoiding the explicit construction of the $n\times n$ correlation matrix. Consider two matrices with the same number of rows, and respective numbers of columns $l_A$ and $l_B$,
\[
    \mat{A}
    =
    \left[
        \vb{a}_1,\dots,\vb{a}_{l_A}
    \right],
    \qquad
    \mat{B}
    =
    \left[
        \vb{b}_1,\dots,\vb{b}_{l_B}
    \right].
\]
\citet{Bishop09} define their modulation product as,
\begin{equation}\label{loc_triangle}
    \mat{A}\triangle\mat{B}
    =
    \left[
        \vb{a}_1\odot\vb{b}_1,
        \dots,
        \vb{a}_1\odot\vb{b}_{l_B},
        \dots,
        \vb{a}_{l_A}\odot\vb{b}_1,
        \dots,
        \vb{a}_{l_A}\odot\vb{b}_{l_B}
    \right],
\end{equation}
\noindent 
where $\mat{A}\triangle\mat{B}$ shares the same number of rows as $\mat{A}$ and $\mat{B}$ but has expanded to $l_A\times l_B$ columns. A shorthand notation for a repeated self-modulation also follows,
\begin{equation}\label{self-modulation}
    \mat{A}^{\! \triangle x} = \underset{1}{\mat{A}}\triangle\underset{2}{\mat{A}}\dots\triangle\underset{x}{\mat{A}},
\end{equation}
\noindent where $\mat{A}^{\! \triangle x}$ contains $\left(l_A\right)^x$ columns.

In the present study, we raise the correlation matrix to the third power, corresponding to a triple-self-modulation of the square root matrix $\mat{W}$,

\begin{equation}\label{loc_cubic_cov}
    \left(\mat{\Omega}^f\right)^{\odot 3}=(\mat{W}^f)^{\triangle 3}
    \left[
        (\mat{W}^f)^{\triangle 3}
    \right]^T,
\end{equation}
\noindent 
which strongly damps weak correlations while retaining the sign and most of the amplitude of the important correlation structures in the smoothed ensemble (see \figs \ref{loc}d, \ref{loc}i and \ref{loc}n). These panels reveal that the elementwise cube has succeeded in isolating the most robust flow-adaptive correlation structures, as evidenced by their proximity to the selected point, their structural alignment with the larger weather system, and their correct movement toward the east over time. 

A naive third-order modulation product would contain one column for every ordered triple of ensemble members, giving \(k^3\) modulated members following \eq \eqref{self-modulation}. For the \(k=24\) member lagged ensemble used here, this would produce \(24^3=13824\) members. However, we note that many of these columns are identical due to the commutativity of the Schur product, meaning it is possible to combine repeated members into a compact square root that reproduces the elementwise cubed covariance exactly. 

In general, the number of unique modulated members $m$ produced by an ensemble of size $k$ when self-modulated $x$ times is
\begin{equation}\label{loc_m}
    m(k,x) = {k+x-1 \choose x},
\end{equation}
\noindent where ${n \choose k} = \frac{n!}{k!\left(n-k\right)!}$ represents the number of unique ways to select $k$ elements from a set of size $n$.

In our case where $x=3$, we wish to index our $m$ unique modulated members in terms of the three original members ($a$, $b$ and $c$) they are comprised of. Let $\mathcal{I}_3(k)$ denote this ordered set of unique index combinations,
\begin{equation}\label{loc_set}
\mathcal{I}_3(k)
=
\left\{
(a,b,c): 1\leq a\leq b\leq c\leq k
\right\},  \quad \left|\mathcal{I}_3(k)\right| = m
\end{equation}

We now define the reduced modulated ensemble $\mat{W}_{\mathrm{loc}}^f\in\mathbb{R}^{n\times m}$ where each of its $m$ unique columns corresponds to one unique index combination from $\mathcal{I}_3(k)$,

\begin{equation}\label{loc_Wloc}
\mat{W}_{\mathrm{loc}}^f
=
\left[
\sqrt{\eta_{abc}}
\left(
\vb{w}^f_a \odot \vb{w}^f_b \odot \vb{w}^f_c
\right), \dots
\right]_{(a,b,c)\in\mathcal{I}_3(k)} ,
\end{equation}
\noindent 
and \(\vb{w}^f_a\) denotes the \(a\)th column of \(\mat{W}^f\). Importantly, unique columns must also be weighted by the square root of the number of times they repeat in the full ordered expansion ($\eta_{abc}$) to ensure their covariance contribution is unchanged. $\eta_{abc}$ depends on the unique index combination $(a,b,c)$ as follows, 

\begin{equation}\label{loc_eta}
    \eta_{abc}
    =
    \begin{cases}
        1, & a=b=c,\\
        3, & \text{two of } a,b,c \text{ are equal},\\
        6, & a,b,c \text{ are all distinct}.
    \end{cases}
\end{equation}

In our case $k = 24$ and $x=3$, and \eq \eqref{loc_m} gives \(m={24+3-1\choose3}=2600\) modulated members, which is an $\sim80\%$ reduction from the \(13824\) members produced by the naive ordered expansion. Note in the original SENCORP implementation \citep{Bishop07}, smoothed correlations were raised to an even power before being modulated with the original (unsmoothed) perturbations to restore the ensemble variance. For the squared correlation case, \eq \eqref{loc_m} indicates this method would result in $k \times m(24,2) = 24\times {24+2-1\choose 2}=7200$ modulated members. We found that this approach offered negligible improvement over the triple-self modulation approach we implement here, despite requiring $\sim$3 times as many ensemble members. The compactness of our localization procedure renders it amenable for practical use on full-sized numerical weather prediction fields.  

\begin{figure}[!ht]
    \centering
    \noindent\includegraphics[width=0.5\textwidth]{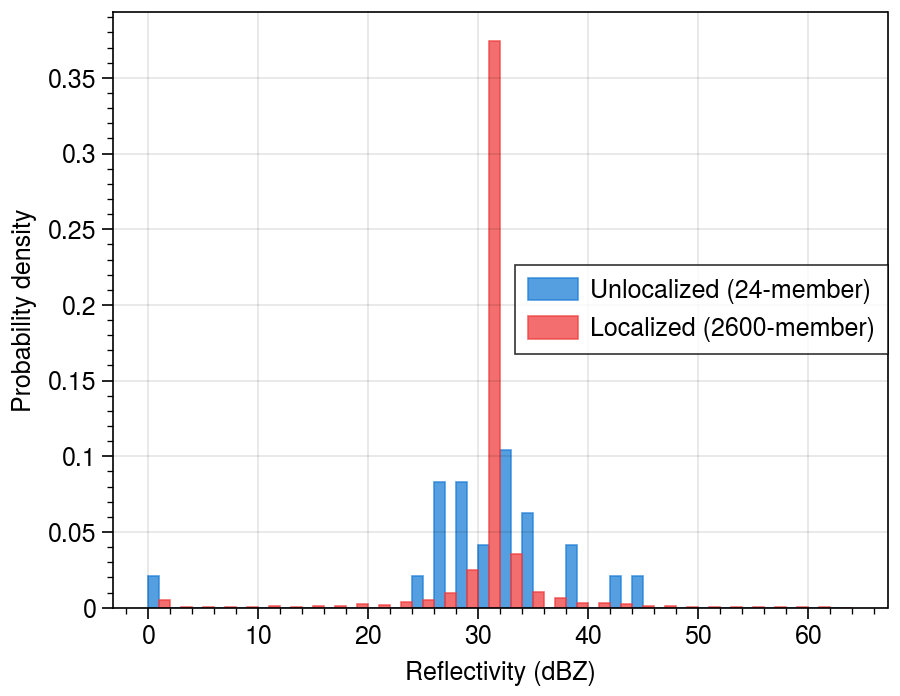}\\
    \caption{The prior reflectivity distribution for $\lambda = 1$ at the initial time (+0 min.) and the selected point in \fig \ref{loc}. Probability density is shown for members from the raw 24-member forecast ensemble (blue) and the localized 2600-member modulated ensemble (red).}\label{prior_distro}
\end{figure}

The remaining task for yielding the localized ensemble perturbations $\mat{Z}^f_{\mathrm{loc}}$ is to prescribe an appropriate variance to correspond to the initial ensemble. This is done by pre-multiplying the perturbation matrix by the square root of the maximum of the smoothed or unsmoothed variances at each grid point, $\bv{\sigma}_\mathrm{max}=\max\left[\bv{\sigma},\bv{\omega}\right]$. Smoothing typically lowers the ensemble variance, hence the unsmoothed variance is typically larger and $\bv{\sigma}_\mathrm{max}\approx\bv{\sigma}$. An exception often occurs near patchy convection when all or a majority of ensemble members miss the precipitation due to sampling bias, producing spuriously low or zero variance. The smoothed variance is often $\omega>\sigma$ at such grid points as smoothing has helped spread the ensemble variance spatially. This formulation reduces the occurrence of situations where ensemble spread is either zero or otherwise spuriously low due to the limited number of ensemble members, permitting a suitable ETKF correction where it would otherwise be ignored. The final localized perturbations are then written as follows:

\begin{equation}\label{loc_Zloc}
    \mat{Z}^f_{\mathrm{loc}}
    =
    \lambda
    \diag\left(\sqrt{\bv{\sigma}_{\mathrm{max}}}\right)
    \mat{W}^f_{\mathrm{loc}},
\end{equation}

where \(\lambda\) is an inflation factor tuned in Section \ref{sec:tuning}. \fig \ref{prior_distro} shows how members from the localized ensemble computed using \eq \eqref{loc_Zloc} compare to those from the raw 24-member ensemble for the single point highlighted in \fig \ref{loc}. The figure shows that aside from damping spurious correlations, localizing using a modulated ensemble also infills and expands the sparse 24-member sample distribution into a smoother 2600-member distribution while reproducing its overall shape and position. These extra degrees of freedom are critical for assimilating dense, highly variable fields such as radar imagery. 

\subsection{Time-Dependent Observation Error}\label{rsched}

The preemptive forecast defined in Section~\ref{pef} applies a single set of analysis weights $\bv{w}$, computed from observations at the analysis time, to the forecast perturbations at every subsequent lead time. This approach relies on the linearity assumption in \eq \eqref{assumption}, meaning errors accumulate over time as the true forecast evolves, leading to progressively poorer analyses at longer lead times. It is therefore desirable for the analysis increments to relax towards the initial ensemble forecast at long lead times, instead of producing increasingly spurious corrections. We achieve this by inflating the observation error with lead time, reflecting the ill-posed nature of our linearity assumption for longer forecasts:

\begin{equation}\label{rsched_s}
    \mat{R}^{-1/2}(t) = s(t)\,\mat{R}^{-1/2}, \qquad s(t) =\exp\!\left(-t/\tau\right) ,
\end{equation}
where the $\mat{R}$ inflation is expressed in terms of its inverse square-root for notational convenience and $\tau$ is a tunable parameter.

A naive implementation of \eq \eqref{rsched_s} for the ETKF would entail recomputing the singular value decomposition in \eq \eqref{svd} at every lead time to reflect the changing $\mat{R}$. However, we note that the time dependence in \eq \eqref{rsched_s} is a single scalar $s(t)$ that factors cleanly through the SVD by simply scaling singular values while leaving singular vectors unchanged. To show this, we substitute $\mat{R}^{-1}(t)=s(t)^2\mat{R}^{-1}$ into the $k\le p$ branch in \eq \eqref{eq:branches-a}, giving
\begin{equation}
\begin{aligned}\label{rsched_branch_a}
    (\mat{H}\mat{Z})^T \mat{R}^{-1}(t)\, \mat{H}\mat{Z}
    &= s(t)^2\,(\mat{H}\mat{Z})^T \mat{R}^{-1} \mat{H}\mat{Z}\\
    &= \mat{C}_k\!\left[s(t)^2\mat{\Gamma}\right]\!\mat{C}_k^T,
\end{aligned}
\end{equation}
so the singular vectors $\mat{C}_k$ are exactly those of \eq \eqref{eq:branches-a} and the singular values become $\mat{\Gamma}(t)=s(t)^2\mat{\Gamma}$. 

Likewise, substituting $\mat{R}^{-1/2}(t)=s(t)\mat{R}^{-1/2}$ into the $k>p$ branch \eq \eqref{eq:branches-b} leaves $\mat{E}$ unchanged and scales its singular values to $\mat{\Lambda}(t)=s(t)^2\mat{\Lambda}$,

\begin{equation}
\begin{aligned}\label{rsched_branch_b}
    \mat{R}^{-1/2}(t)\mat{H}\mat{Z}\left(\mat{R}^{-1/2}(t)\mat{H}\mat{Z}\right)^T
    &= s(t)^2\,\mat{R}^{-1/2}\mat{H}\mat{Z}\left(\mat{R}^{-1/2}\mat{H}\mat{Z}\right)^T\\
    &= \mat{E}\!\left[s(t)^2\mat{\Lambda}\right]\!\mat{E}^T,
\end{aligned}
\end{equation}

while $\mat{C}_p$ of \eq \eqref{eq:branches-c} is itself invariant, because the two $s(t)$ factors cancel,

\begin{equation}
\begin{aligned}\label{rsched_branch_c}
    \mat{C}_p(t)
    &= \left(\mat{R}^{-1/2}(t)\mat{H}\mat{Z}\right)^T\mat{E}\,\mat{\Lambda}(t)^{-1/2}\\
    &= s(t)\left(\mat{R}^{-1/2}\mat{H}\mat{Z}\right)^T\mat{E}\,
      \left[s(t)^2\mat{\Lambda}\right]^{-1/2}\\
    &= \mat{C}_p .
\end{aligned}
\end{equation}

In both ETKF branches, the singular vectors ($\mat{C}_k$, or $\mat{E}$ and $\mat{C}_p$) are independent of $s(t)$. A single decomposition computed with the initial $\mat{R}$ therefore calculates the singular vectors once, and the update at each lead time is recovered cheaply by rescaling the singular values according to the time-dependence, $\mat{\Gamma}(t)=s(t)^2\mat{\Gamma}$ or $\mat{\Lambda}(t)=s(t)^2\mat{\Lambda}$. Scaled singular values are calculated separately for each time-block and substituted into \eqs \eqref{etkf_mean} \& \eqref{etkf_pert} to retrieve the analysis mean and perturbations, respectively.

The schedule interpolates between two limits. As $\tau\!\to\!\infty$, $s(t)\!\to\!1$ at all times and the standard preemptive update of Section \ref{pef} is recovered exactly. As $s(t)\!\to\!0$, the singular values $\mat{\Gamma}(t)\!\to\!0$, so the mean increment in \eq \eqref{etkf_mean} vanishes and the perturbation transform in \eq \eqref{etkf_pert} approaches the identity: the analysis relaxes smoothly back to the prior mean and the prior spread. A finite $\tau$ thus places greatest trust in the observations near the analysis time, where the linear assumption holds best, and progressively defers to the model forecast at longer lead times.

\subsection{Model and Radar Data}\label{model}

\begin{figure}[!ht]
    \centering
    \noindent\includegraphics[width=0.5\textwidth]{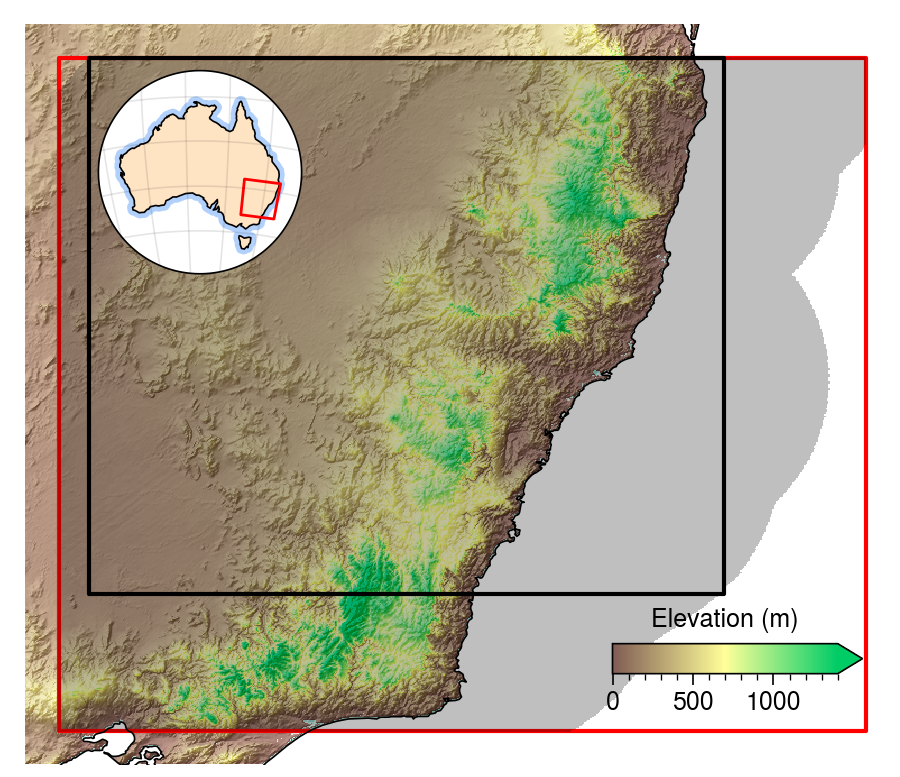}\\
    \caption{Model and radar data domain. The ACCESS-CE convection-allowing model domain (red box; 144–156°E, 28–38°S) centered on Sydney in southeastern Australia, shaded by terrain elevation (m). The subset domain used in this study (black box; 144.46-153.88°E, 28.00-35.96°S), lies completely within the bounds of the multi-radar mosaic observations (grey shading).}\label{fig:model}
\end{figure}

Ensemble forecasts are accessed from a historical archive of the Bureau of Meteorology's operational ACCESS City model over the Sydney domain (ACCESS-CE, see \fig \ref{fig:model}), the convection-permitting limited-area ensemble of ACCESS-C \citep{Rennie22, Cooper25}. The APS3 version of ACCESS-CE uses the Met Office Unified Model version 10.6 with the ENDGame dynamical core and explicitly resolved deep convection. The Sydney domain uses the mid-latitude RAL1-M physics configuration, has 80 vertical levels, and is run on a variable-resolution grid with an inner horizontal grid spacing of approximately 2.2 km. Operational ACCESS-CE forecasts contain 12 members per cycle and are initialized once every 6 hours. Importantly, radar reflectivity is not directly assimilated in the operational 4DVAR DA scheme due to implementation difficulties \citep{Rennie20}. Instead, model precipitation is updated using latent heat nudging \citep[LHN;][]{Jones97} for the first hour of the forecast (the DA window) via a 15-minute rainfall accumulation product \citep{Seed07}. Anecdotal experience from forecasters and model evaluation \citep{Cooper25} suggest that precipitation features are frequently misplaced due to a lack of high-resolution radar assimilation, and the associated uncertainties are underestimated due to an under-spread ensemble. This status creates an operational need to adjust the current model guidance based on recent radar observations, exactly the aim of our preemptive forecasts.

In this study, we form a 24-member lagged ensemble by concatenating the 12 members from each of the two most recent ACCESS-CE cycles. The forecast variable used is simulated radar reflectivity at 1 km above ground level, calculated within the Unified Model using the forward operator outlined in \citet{Stein14}. Reflectivity is expressed in dBZ, and values below 0 dBZ are clipped for both the model and observations to match the effective sensitivity of the operational radar mosaic. The study domain is restricted to the part of ACCESS-CE Sydney domain covered by the multi-radar reflectivity mosaic (see \fig \ref{fig:model}), yielding a 476 $\times$ 402 grid over southeastern Australia. The radar mosaic is a ``ground level'' reflectivity product, calculated using the lowest available radar scans and subsequently re-gridded to the ACCESS-CE grid using xESMF \citep{Zhuang25}. This results in a height mismatch from the modeled fields at 1 km altitude as the lowest radar observation heights vary with distance. This, along with other quality control issues such as bright band and clutter contamination, will be addressed by tuning the observation error covariance.

\subsection{Verification methods}
\label{sec:verification}

The primary aim of this paper is to introduce the methodology required to make skillful preemptive forecasts of operationally important variables, in this case radar reflectivity (Sections~\ref{etkf}--\ref{rsched}). Here, we offer a brief description of the probabilistic verification used to demonstrate this added skill. The field of probabilistic rainfall verification is a vast and active area of research, and as such we do not intend this verification to be exhaustive. Rather, we present only the most common verification metrics, including point-based categorical scores, probabilistic fractions skill scores, Brier scores, and reliability diagrams. These metrics are all briefly introduced below, but please refer to the cited literature for more details. Further, the preemptive forecasting methodology is a general forecast-correction technique grounded in ensemble data assimilation theory. Its potential use cases are also general, including any number of observation sources or forecast variables. The radar forecasting application presented here represents a single, particularly challenging example of all such potential use cases. Therefore, we do not aim to benchmark our findings for this specific use case, and acknowledge the existence of a range of specialist radar nowcasting techniques (either model-based \citep{Heinselman24}, statistical \citep[\eg][]{Pulkkinen19}, or machine-learned \citep{Ravuri21}). Rather, we aim to demonstrate an improvement relative to the background NWP forecasts described in Section~\ref{model}. 

\begin{table}[!ht]
\caption{Reflectivity thresholds used for verification, with the equivalent rain rate from the Marshall--Palmer relation ($Z = 200\,R^{1.6}$) and the conventional rainfall-intensity class.}\label{t:zr}
\centering
\begin{tabular}{lrl}
\toprule
Reflectivity (dBZ) & Rain rate (mm\,hr$^{-1}$) & Description\\
\midrule
10 & 0.15 & drizzle \\
20 & 0.65 & light rain \\
30 & 2.73 & moderate rain \\
\bottomrule
\end{tabular}
\end{table}

The verification employed here follows the common practice of evaluating forecasts at various operationally significant reflectivity/rainfall thresholds, which are summarized in Table~\ref{t:zr}. Common point-wise metrics are included, but our verification focuses mostly on neighborhood matching approaches to avoid ``double penalty'' biases \citep{Ebert08}. This approach acknowledges that forecasts on the convective scale may contain small displacement errors while remaining meteorologically useful. All neighborhood probabilities follow the neighborhood ensemble probability (NEP), or ``smoothing'', formulation of \citet{Schwartz17}, where grid point ensemble exceedance probability is averaged over a square neighborhood, with the verifying observations treated consistently as the neighborhood observed event fraction over the same window. The subset of observations that were assimilated at the analysis time are omitted from verification scores. The neighborhood-maximum (``searching'', NMEP) variant is not used due to the impracticality of searching neighborhoods for each of the 2600 members in our modulated ensemble. This also precludes useful verification of the high thresholds (\eg $\geq$40 dBZ) that describe rare, extreme weather events such as hailstorms, because NEP-based metrics suffer severely from the neighborhood averaging of small-base-rate events, causing a rapid loss of sharpness\footnote{``Sharpness'' refers to the ability to produce confident probabilistic estimates close to 0 or 1 \citep{Murphy93}.}. For this reason, we also do not calculate the neighborhood continuous ranked probability score (CRPS), because equal weight is given to operationally unimportant thresholds below 10 dBZ and to poorly verified thresholds above 40 dBZ. Verifying more extreme, high-impact events using suitable methodologies will be a subject of future work \citep[\eg object-based verification;][]{Skinner18}.

\subsubsection{Point-wise/Categorical scores}
\label{sec:categorical}

The simplest metric used in this paper is a direct point-based, root mean square error (RMSE) against radar observations pooled over all ensemble members. Pointwise categorical skill is also assessed for threshold exceedance using contingency tables containing hits ($a$), false alarms ($b$), misses ($c$), and correct negatives ($d$). From these counts, we compute the probability of detection, $\mathrm{POD}=a/(a+c)$, the success ratio $\mathrm{SR}=a/(a+b)$, (equivalently one minus the false alarm ratio, 1- FAR), the critical success index, $\mathrm{CSI}=a/(a+b+c)$, and the frequency bias, $\mathrm{FB}=(a+b)/(a+c)$ \citep{Schaefer90,Jolliffe12}. Rather than verifying the ensemble mean for the categorical metrics, we accumulate and pool over all ensemble members, grid points, and cases before forming the ratios. These scores therefore summarize the typical single-member categorical skill of the ensemble. The four quantities are displayed jointly on a performance diagram \citep{Roebber09}, on which CSI and frequency bias appear as reference contours, making the trade-off between detection and false alarms directly visible across thresholds and lead times.

\subsubsection{Probabilistic fractions skill score}

The fractions skill score (FSS) is a displacement-tolerant spatial skill metric produced by comparing neighborhood fractions of forecast and observed threshold exceedance \citep{Roberts08, Ebert08}. Here, we use the probabilistic version of the FSS \citep[pFSS;][]{Necker24}, computed by combining the neighborhood and ensemble probabilities following the \citet{Schwartz17} NEP formulation. Forecast and observed fractions are computed over the same valid footprint at each neighborhood, so that missing-radar or assimilated grid points do not bias the comparison. The pFSS is computed over multiple reflectivity thresholds and neighborhood sizes. 

\subsubsection{Neighborhood Brier score and Brier skill score}

The neighborhood Brier score (BS) represents the mean squared difference between the NEP and the observed neighborhood fraction, with lower values indicating closer probabilistic agreement \citep{Brier50,Gneiting07,Wilks11,Jolliffe12,Schwartz17}. It is therefore the probabilistic analog of a mean squared error, and its integration over all possible thresholds gives the CRPS \citep{Gneiting07}. The Brier skill score is formed relative to a climatological neighborhood forecast as $\mathrm{BSS}=1-\mathrm{BS}/\mathrm{BS}_{\mathrm{ref}}$, where $\mathrm{BS}_{\mathrm{ref}}$ is the reference Brier score obtained by replacing the forecast NEP with the observed climatological neighborhood fraction for that threshold and window, evaluated over the full valid domain. Positive values therefore indicate improvement over the climatological neighborhood probability forecast.

\subsubsection{Reliability diagrams}

Reliability diagrams assess the calibration of NEPs against observed neighborhood fractions. Here, calibration refers to the agreement between forecast probability and the conditional mean observed fraction: for example, forecasts with a NEP of $0.5$ should, on average, have an observed neighborhood fraction close to $0.5$. NEPs are grouped into probability bins, and the mean forecast NEP in each bin is compared with the mean of the corresponding observed neighborhood fractions \citep{Murphy73,Wilks11,Jolliffe12,Schwartz17}. A well-calibrated forecast lies close to the one-to-one line, while curves below or above the diagonal indicate overforecasting or underforecasting, respectively. A normalized forecast-frequency histogram is also shown beneath each diagram to indicate the sharpness of the forecast probabilities and identify poorly populated bins that should not be over-interpreted.

\subsubsection{Statistical significance}
\label{sec:significance}

To establish the robustness of all metrics and to determine whether the difference between the preemptive forecasts (PEF) and the NWP baseline is statistically significant, we follow \citet{Schwartz17} in performing significance testing via paired bootstrap analysis \citep{Hamill99,Wolff14}. A $99\%$ confidence interval is formed for all metrics and PEF-NWP (paired) differences by resampling with replacement and re-accumulating the statistics 1000 times. Differences in quantities are deemed statistically significant if the resulting paired-bootstrap confidence interval does not overlap 0. Note that our nine-week verification dataset contains individual cases 5 hours apart, and these may not be fully independent from one another. This temporal dependence mean the estimated 99\% confidence intervals may be overstated.

\subsection{Model Tuning}
\label{sec:tuning}

There remain several implementation choices and tunable parameters to be finalized in our preemptive forecasting system. First, we make the simplifying assumption that the observation error covariance is a scalar, diagonal matrix $\mat{R} = R \cdot \mat{I}_p$. This is also implemented and tuned in terms of its inverse square root for notational clarity (the inverse observation-error standard deviation, $R^{-1/2}$), and avoids the matrix inversions that would otherwise be required in \eg \eq \eqref{svd} for a full $\mat{R}$. 

Next, seeing as our forecast and observations contain a common variable on a common grid, our observation operator $\mat{H}$ becomes a grid-point selection of observations coincident with the initial forecast time\footnote{We ensure the initial PEF time (\ie the analysis time), is no shorter than two hours after the base time of the NWP forecasts used. This imposes the practical constraint of not having immediate access to the forecast due to the ACCESS-CE runtime.}. A single 2D radar observation field contains $476\times402=191352$ grid points, which is two orders of magnitude larger than the number of members in our modulated ensemble. Recall that the number of ensemble members places an upper limit on the number of independent observational directions in the ensemble subspace \citep[\eg][]{Evensen03}. Therefore, assimilating every available observation with the ETKF is both ineffective and computationally expensive. We therefore define a subroutine that subsamples $p$ points from the radar image by selecting points more densely in regions with relatively larger ensemble variance and none in regions with zero coincident ensemble spread. We found the forecasts were broadly insensitive to the specifics of this observation subselection procedure and omit further details for brevity.

There are five tunable parameters defined to this point: the inverse observation-error standard deviation $R^{-1/2}$, the number of assimilated observations $p$, the localization smoothing length $\ell$, the ensemble inflation $\lambda$, and the lead-time decay scale $\tau$. Tuning these parameters is a vital step in producing skillful forecasts and takes place using a Bayesian hyperparameter optimization software package \citep[Optuna;][]{Akiba19}. The package evaluates the suitability of parameters by performing repeated trials with a user-defined objective function ($J$) that should reward skillful performance. Here, we define the objective in terms of the skill improvement of the preemptive forecast (PEF) relative to the NWP, pooled over our predefined rainfall thresholds $q\in\{10,20,30\}\,\mathrm{dBZ}$, neighborhood windows $L\in\{10,25\}$ grid cells, forecast lead times $t\in\{0,2,6\}$ hours, and training cases ($j$), 

\begin{equation}\label{J}
J = \sum_{i\in(q,L,t,j)} \mathcal{L}(\mathrm{BS}^{\mathrm{PEF}}_i - \mathrm{BS}^{\mathrm{NWP}}_i),
\quad
\mathcal{L}(x) =
\begin{cases}
x, & x \le 0,\\
\beta\,x, & x > 0.
\end{cases}
\end{equation}

\begin{figure*}[ht]
    \centering
    \noindent\includegraphics[width=\textwidth]{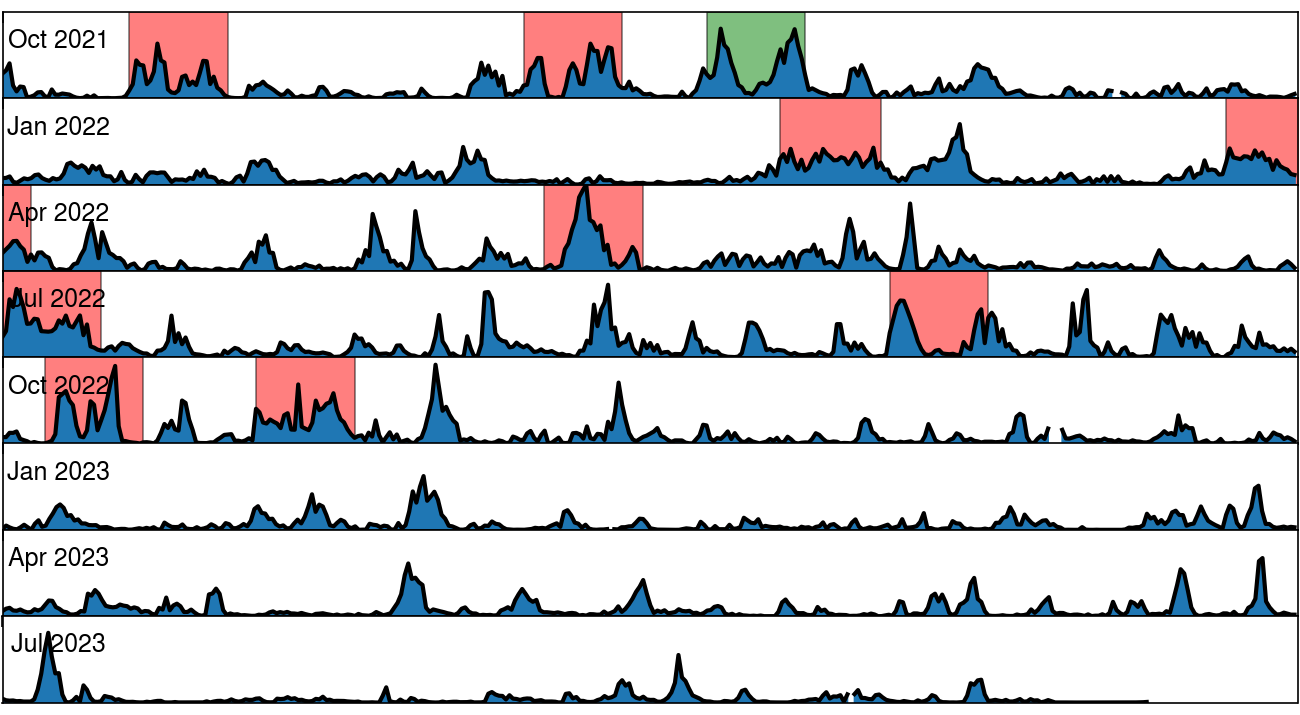}\\
    \caption{Time series of accumulated model-derived reflectivity (clipped below 0 dBZ, then summed) over a two-year ACCESS-CE Sydney archive period. Tall peaks indicate periods with significant rainfall, and nine highly active seven-day trial windows were selected on this basis for verification. Verification weeks are highlighted in red and the parameter-tuning week is highlighted in green.}\label{training}
\end{figure*}

We make the subjective choice of adding an asymmetric weight $\beta = 3$, which penalizes skill degradation more strongly than it rewards skill improvement. This encourages the parameters to ``do no harm'' by avoiding over-confident, spurious preemptive forecasts that degrade the performance of the initial NWP forecast at longer lead times. Without this asymmetric weighting, we found that parameters were tuning toward producing larger improvements at the initial times at the cost of degrading the long lead times relative to the NWP. Lower $J$ is better, and $J<0$ certifies that the analysis is, on aggregate, more skillful than the NWP forecast. Training takes place using one of ten active weeks selected from two years of archived cases (one week contains 33 individual cases, one every 5 hours) and is highlighted with green in \fig \ref{training}. A single week was subjectively chosen for training due to the significant computational cost of the Bayesian hyperparameter optimization. All cases from the training week were held out of the verification dataset (which is highlighted red in \fig \ref{training}), and training on a different week showed that the forecast verification results changed little (not shown). The tuned parameters were applied unchanged to the remaining nine verification weeks (294 cases), which represent a sample of all four seasons while remaining small enough to be computationally manageable.

\section{Qualitative Evaluation}\label{results}

The parameters selected by the tuning procedure are listed in Table~\ref{t:params} and warrant some interpretation. Tuning led to the inclusion of $p=9000$ observations, a factor of $\sim$3.5 larger than the 2600-member ensemble size. This is an expected result for practical applications where dense radar observations may not be entirely independent and the effective local observation dimension\footnote{The effective local observation dimension is the number of observations that effectively contribute to a single analysis after accounting for localization weights \citep{Kirchgessner14}.} may be much less than the raw number of observations due to the localization scheme \citep{Kirchgessner14}. The observation errors are tuned high ($R^{-1/2}=0.09$ corresponds to a standard deviation of $\sim$11 dBZ), meaning a single observation is trusted only moderately, reflecting representativeness errors due to height mismatches and other quality-control issues noted for the radar observations in Section~\ref{model}. The low-pass filter cutoff used in the flow-adaptive localization is suitably strict at $\ell\approx10$ grid cells, meaning only the broadest scales enter the smoothed perturbations to offset the spectral narrowing induced by the cubic modulation. The posterior spread is also substantially inflated by a factor $\lambda\approx3.0$ to account for the common symptom of ETKF overconfidence with large numbers of observations \citep{Hotta21} and the noted under-dispersion of the prior ensemble \citep[by a factor of $\sim$2 for some measures;][]{Cooper25}. Finally, perhaps the most instructive parameter is the observation-trust decay scale $\tau\approx136\,$min. This formulation damps ETKF singular values to 41.4\%, 17.1\%, 7.1\%, and 2.9\% of their original values after 1, 2, 3, and 4 hrs, respectively. The schedule therefore strongly attenuates corrections at longer lead times, allowing the analysis to relax back toward the prior forecast. This reflects the unsuitability of our linear, preemptive corrections at long lead times given the complex evolution of radar reflectivity fields. Bayesian hyperparameter tuning resulted in a negative objective value of $J=-0.0146<0$ after 452 trials, which confirms preemptive forecasts were on average more skillful than the NWP priors within the training window. 

\begin{table}[!ht]
\caption{Trained pre-emptive forecast parameters selected by minimizing the skill objective $J = -0.0146$ (Section~\ref{sec:tuning}).}\label{t:params}
\centering
\begin{tabular}{llr}
\toprule
Parameter & Description & Value\\
\midrule
$R^{-1/2}$ & Inverse observation error (dBZ$^{-1}$) & 0.09 \\
$p$ & Number of assimilated observations & 9000 \\
$\ell$ & Localization smoothing cutoff (grid cells) & 10.06 \\
$\lambda$ & Posterior inflation factor (unitless) & 2.99 \\
$\tau$ & Observation-trust decay scale (min.) & 136 \\
\bottomrule
\end{tabular}
\end{table}

\begin{figure*}[ht]
    \centering
    \noindent\includegraphics[width=\textwidth]{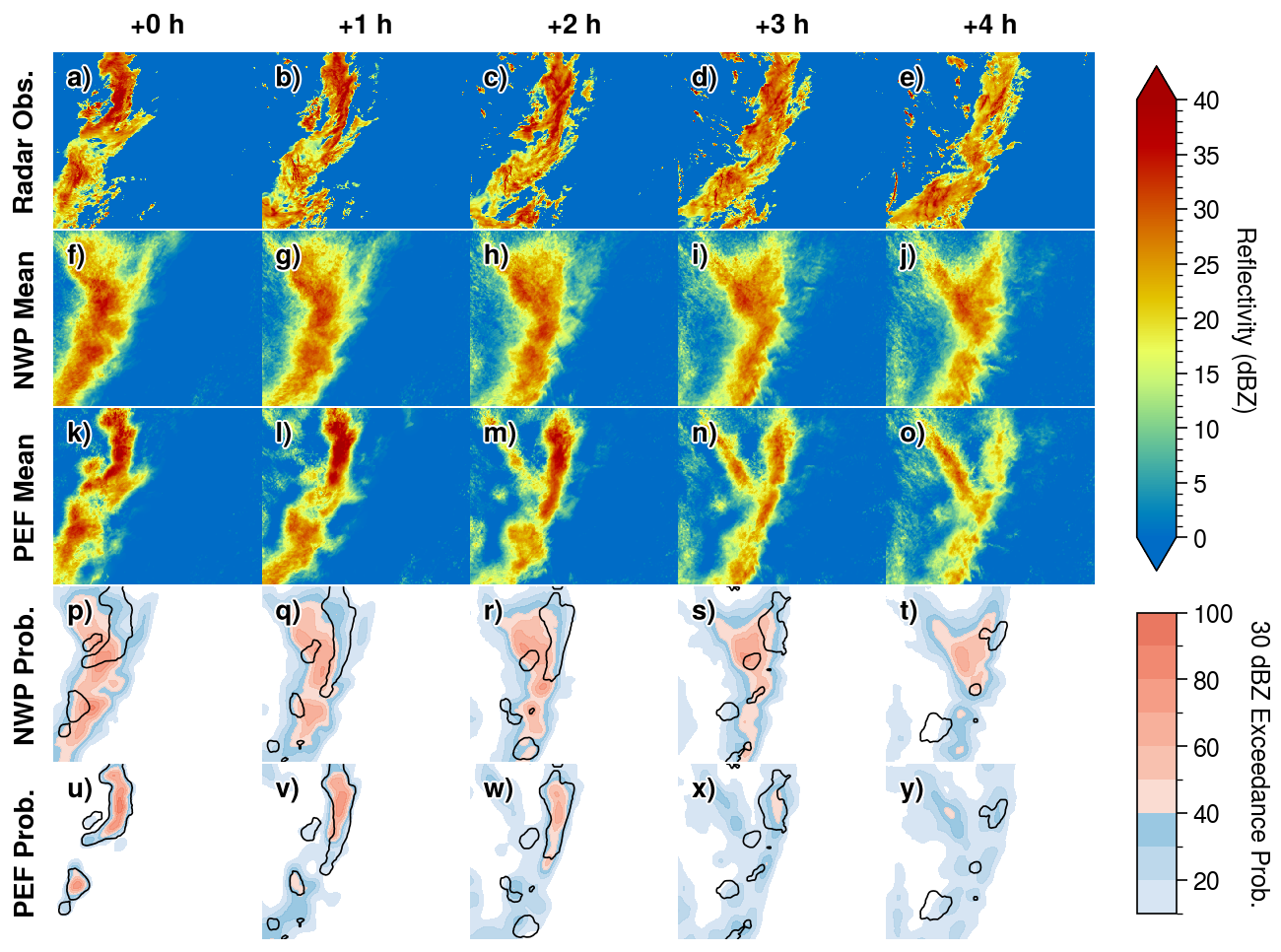}\\
    \caption{Example case 1. Reflectivity and $30\,$dBZ exceedance probability at lead times $0$–$4\,$h. Rows contain radar observations, the NWP forecast ensemble mean, the PEF analysis ensemble mean, the NWP neighborhood exceedance probability, and the PEF neighborhood exceedance probability (both 25-cell neighborhood windows). Black lines show the 20\% contour of the observed neighborhood fraction. The PEF analysis repositions the linear feature eastward based on the observations and maintains this correction throughout the 4-hour forecast window.}\label{example1}
\end{figure*}

\begin{figure*}[ht]
    \centering
    \noindent\includegraphics[width=\textwidth]{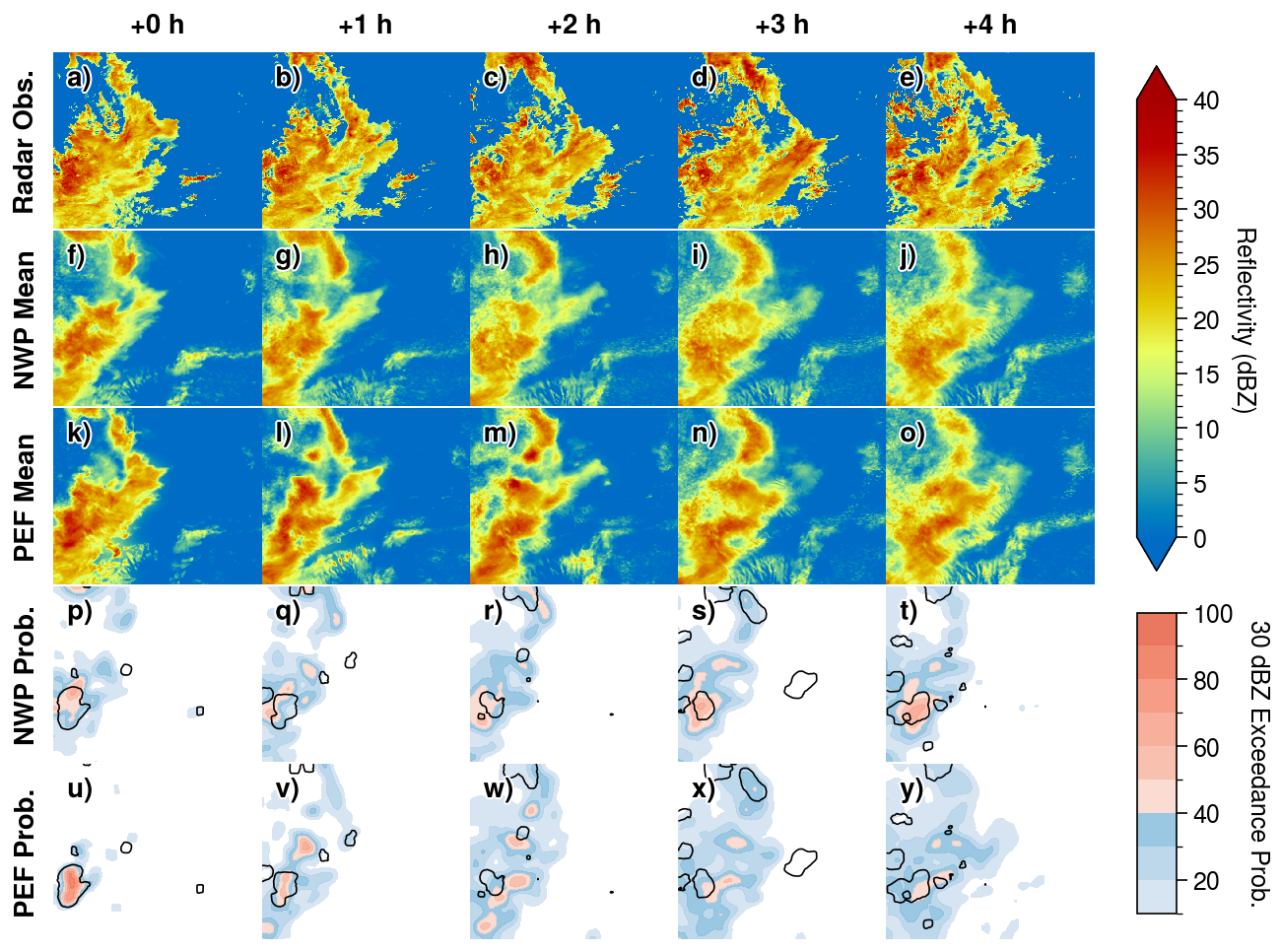}\\
    \caption{Example case 2, see \fig \ref{example1} for description. This case exhibits more complex, nonlinear evolution of the precipitation feature over time. Despite faithfully reproducing the system at the analysis time, the preemptive forecast improvements persist weakly through time and do not outperform the NWP after 1 hour.}\label{example2}
\end{figure*}

We now introduce two example cases drawn from the verification period to develop an intuition of how preemptive forecasts alter their NWP priors. The first (0100 UTC 8 Sep 2022, Fig.~\ref{example1}) demonstrates a near-ideal case for preemptive forecasting and the second (2000 UTC 4 Oct 2022, Fig.~\ref{example2}) is one in which the technique shows little benefit. In both cases the top row shows the radar observations, followed by the NWP and PEF ensemble means. The bottom two rows show the corresponding 30 dBZ neighborhood exceedance probabilities, along with the 0.2 observed neighborhood fraction contour overlaid in black. The 30 dBZ threshold is chosen for illustration due to its operational relevance in being associated with the highest rainfall rates. The neighborhood probability rows illustrate quantities that are used for probabilistic evaluation, including the neighborhood Brier scores employed for model tuning.

The first is a frontal case with a strong linear convective system oriented NNE--SSW. At the analysis time, the raw NWP ensemble (Fig.~\ref{example1}f) confidently places the band too far west, with resulting 30 dBZ probabilities > 50\% running N-- S for roughly the length of the domain (Fig.~\ref{example1}p). This prediction lies well to the west of the main observed contour, and the resulting probability overlap with observations is poor. The preemptive analysis repositions the band eastward and sharpens the ensemble mean and neighborhood probability (Fig.~\ref{example1}k,u) into close agreement with the observations. Crucially, this displacement correction occurs to a coherent, slowly evolving feature that covaries strongly with subsequent time steps (\fig \ref{loc}e, \ref{loc}j, and \ref{loc}o). As a result, the preemptive correction persists through time, and the eastward shift of the precipitation band remains evident through much of the four-hour window (Fig.~\ref{example1}k-o). Aside from the main linear feature, an additional NW--SE aligned feature is correctly suppressed by the PEF in the first 2 hours of the forecast, before spuriously intensifying as the PEF reverts to the NWP at longer lead times. Overall, the PEF neighborhood ensemble probability field tracks the observed contour more closely than the NWP throughout, leading to improved probabilistic skill scores.

The second case is a large, multicellular convective mass undergoing complex evolution along with the formation and decay of isolated cells across the domain (\figs \ref{example2}a-e). Here the PEF faithfully corrects the NWP to match observations at the initial time (\figs \ref{example2}k and \ref{example2}u), but the improvement persists only modestly to the first lead time (+1 hour, \fig \ref{example2}v). The prior ensemble has failed to produce robust correlations through time due to the chaotic, small scale variability in this case, and what corrections are produced by the PEF at +2-3 hours (\eg \figs \ref{example2}w-x) do not obviously align more closely to the observations, indicating the ill-posedness of our linear correction assumption. Once again, the PEF largely reverts to the NWP forecast at later lead times (Fig.~\ref{example2}n,o), as was intended by our observation error inflation scheme. 

\begin{figure*}[ht]
    \centering
    \noindent\includegraphics[width=\textwidth]{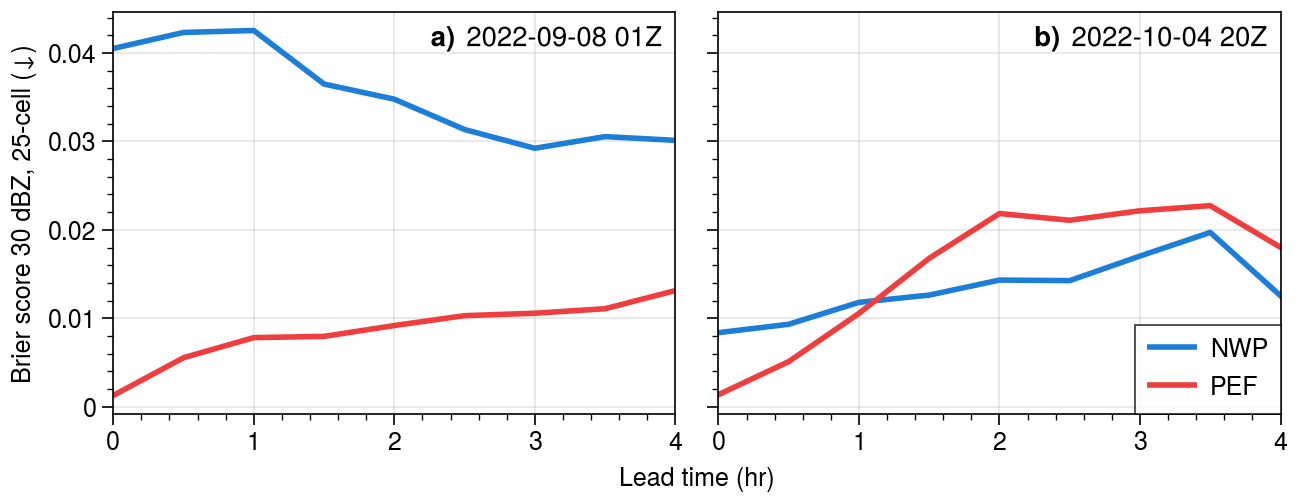}\\
    \caption{Brier score for the two example cases in Figs.~\ref{example1} (panel a) and \ref{example2} (panel b). The neighborhood Brier score (30 dBZ threshold, 25-cell window, lower is better) is displayed versus lead time over a four-hour window for the NWP forecast (blue) and PEF analysis (red).}\label{example_brier}
\end{figure*}

We now find it instructive to see how these contrasting examples performed on the main verification/tuning metric, the Brier score (at thresholds $30$ dBZ, $25$-cell window, illustrated in \figs \ref{example1} \& \ref{example2}). Fig.~\ref{example_brier}a shows that the linear case is near-perfect at the initial time ($\mathrm{BS}\approx0.001$ \vs $\approx0.040$ for the NWP), and, while the PEF score degrades gradually with lead time as expected, it remains well below the NWP curve for the entire four-hour window due to the long-lasting displacement correction. In the strongly nonlinear case (Fig.~\ref{example_brier}b), the analysis again begins close to perfect, but decays more rapidly to $\mathrm{BS}\approx0.02$ at +2 hours due to the spurious corrections produced in the 30 dBZ probability forecasts. The NWP forecast is itself far more skillful for this case, leaving little room for improvement by the PEF. As a result, the PEF's improvement is lost after roughly 1 hour, and remains marginally poorer than the NWP throughout the rest of the forecast period. This type of result is strongly penalized by the asymmetric cost function in \eq \eqref{J} and provides an insight into why a conservative observation precision and error inflation schedule was preferred by the tuning procedure. This has ensured that even for an unfavorable case such as this one, the long-lead degradations remain modest. To summarize, the preemptive forecasts deliver large and durable gains when the NWP forecast can be easily corrected, and the smallest gains when the NWP is already competitive and/or the forecast evolution is unpredictable. We now offer a more comprehensive suite of evaluations to determine whether these qualitative assessments are robust over the remainder of the verification period.

\section{Quantitative Evaluation}\label{results2}

\begin{figure*}[ht]
    \centering
    \noindent\includegraphics[width=\textwidth]{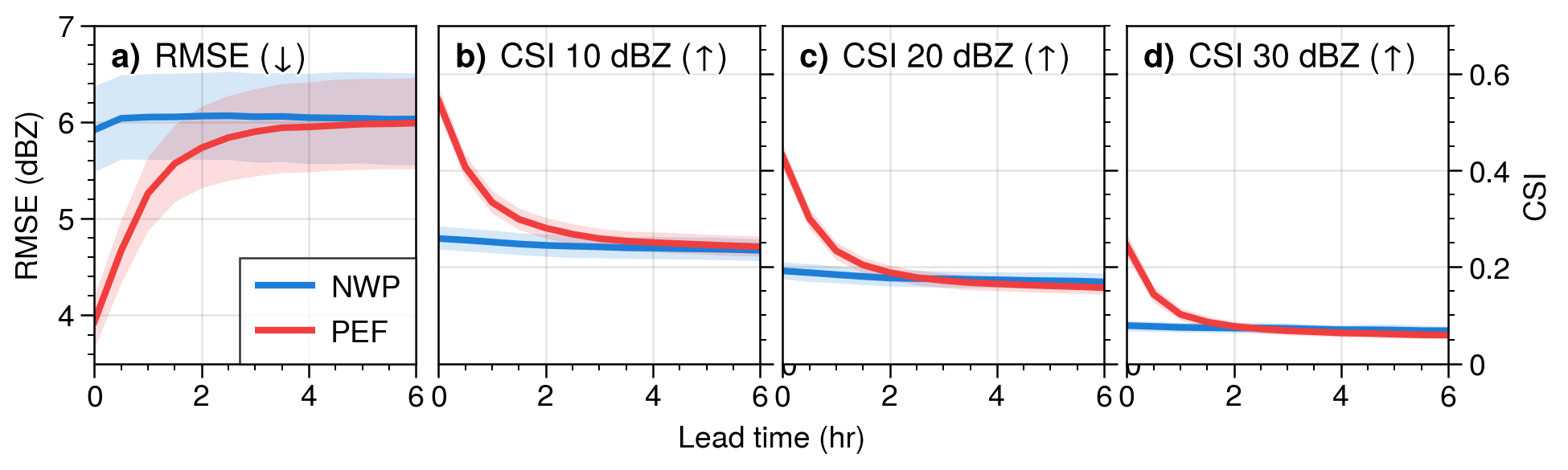}\\
    \caption{Point-based skill versus lead time pooled over all ensemble members and all cases in the 9-week trial period. (a) Reflectivity RMSE (lower is better), and (b--d) critical success index (CSI, higher is better) at 10, 20 and 30 dBZ thresholds. Shaded bands indicate 99\% confidence intervals for NWP forecasts (blue) and PEF analyses (red). The PEF shows improvements on all metrics at short lead times.}\label{rmse}
\end{figure*}

Having developed an intuition from two illustrative cases, we now verify the preemptive forecast against all 294 cases in the 9-week verification period. We proceed from simple, point-based scores that penalize spatial displacement harshly \citep{Roberts08,Ebert08}, to the neighborhood-based, probabilistic scores that the system was tuned for. Figure~\ref{rmse}a shows the point-based RMSE pooled over all ensemble members is significantly improved at the analysis time ($\sim 4$dBZ) relative to the baseline NWP ($\sim 6$dBZ). A narrowing of the 99\% confidence interval at shorter lead times also shows the improvement is robust across the verification dataset despite clear climatological variability in the NWP baseline. Significant RMSE improvements persist for roughly 3 hours before the PEF mean and spread revert back to the respective NWP values after 6 hours. \figs \ref{rmse}b-d also confirm that analysis improvements have been achieved across each of the thresholds of interest (10, 20, \& 30 dBZ) according to the Critical Success Index (CSI). Fractional improvements are largest for the strongest reflectivities, where the $30$ dBZ CSI increases threefold from $\sim$0.08 to $\sim$0.25. PEF scores revert to the NWP after roughly 2-3 hours and decay fastest at the higher thresholds where point-scale predictability is shortest-lived. Once again, and despite the sharper PEF mean having greater exposure to double-penalty errors \citep{Ebert08,Mittermaier14}, point-based PEF scores do not fall appreciably below the NWP at longer lead times.

\begin{figure*}[ht]
    \centering
    \noindent\includegraphics[width=\textwidth]{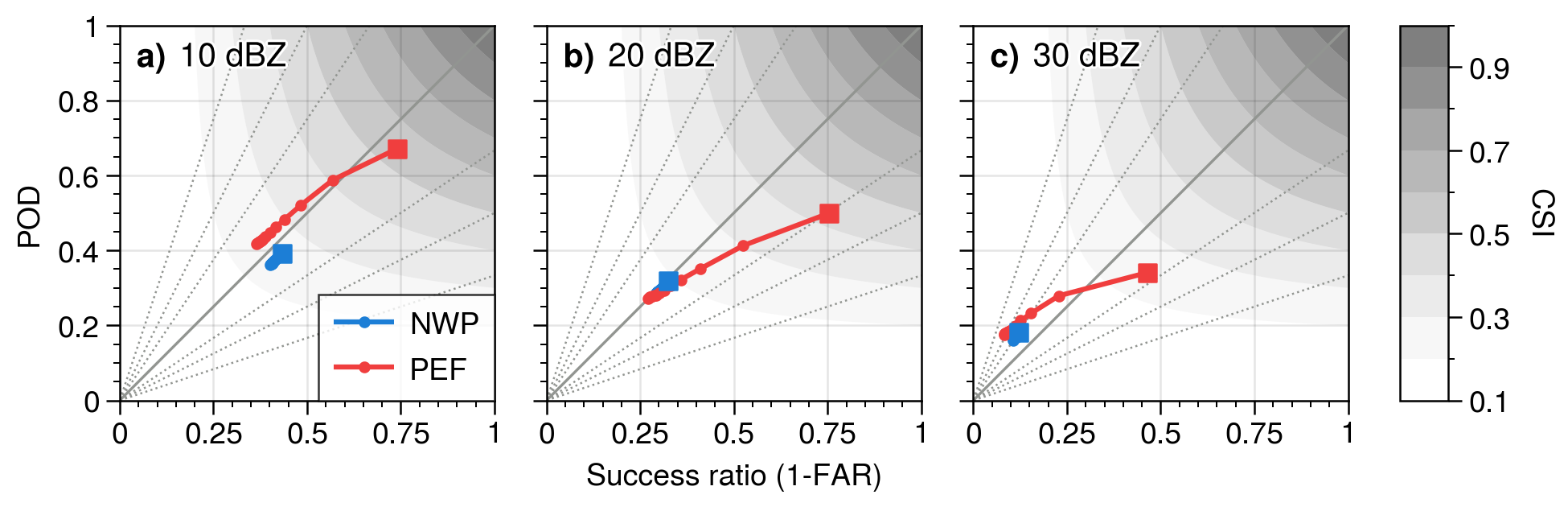}\\
    \caption{Performance diagram \citep{Roebber09} at $10$, $20$ and $30\,$dBZ. Success ratio ($1-\mathrm{FAR}$) versus probability of detection; grey shading is the critical success index and dotted rays the frequency bias. NWP forecast (blue), PEF analysis (red); squares mark the analysis time ($+0\,$h), and lines trace increasing lead time. The PEF analysis begins at markedly higher skill and relaxes toward the NWP forecast with lead.}\label{roebber}
\end{figure*}

\begin{figure*}[ht]
    \centering
    \noindent\includegraphics[width=\textwidth]{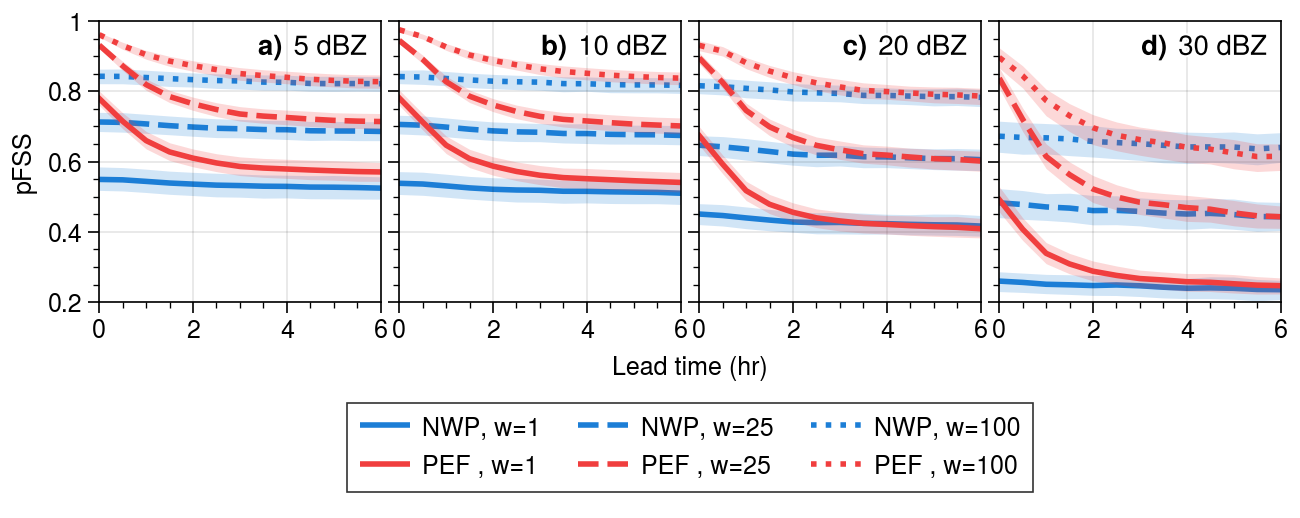}\\
    \caption{Probabilistic fractions skill score (pFSS) versus lead time at 5, 10, 20 and 30 dBZ, for neighborhood windows of 1, 25, and 100 grid cells (solid, dashed and dotted lines, respectively). Shaded bands indicate 99\% confidence intervals for NWP forecasts (blue), PEF analyses (red). Higher is better; skill increases with neighborhood size.}\label{pfss}
\end{figure*}

The performance diagram \citep[\fig \ref{roebber};][]{Roebber09} decomposes the categorical skill score by plotting each lead time as a function of its probability of detection (POD) and success ratio. The PEF analysis shows that the $+0\,$h (squares) sit above and to the right of their NWP counterparts, owing to their higher overall CSI scores. PEF scores do curve marginally below the optimal 45$^\circ$ improvement direction, indicating that increased skill is reducing false alarms more than increasing true detections. However, this apparent underforecasting bias likely arises as a result of tuning the system according to neighborhood metrics where we are rewarded for exceeding these thresholds anywhere within a window, rather than exactly at a point. For all three thresholds, early lead time skill improvements revert back toward the cluster of NWP values, which display a uniform skill and frequency bias throughout the forecast period consistent with Fig.~\ref{rmse}. The uniformity of forecast skill across the first 6 hours of the background NWP model was also noted in \citet{Cooper25} and indicates that very little convective-scale information is imparted by its 4DVAR data assimilation scheme.

Next, Figure~\ref{pfss} evaluates neighborhood skill with the probabilistic fractions skill score (pFSS), which rewards the correct fraction of points exceeding the threshold within a window even if individual pixels are displaced \citep{Roberts08,Schwartz17}. Here, we show the pFSS from neighborhood sizes of 1 (point-wise) to 100 grid points ($\sim$220 km), and observe that while the absolute values change according to the window size, the relationship between the PEF and NWP is qualitatively similar for all\footnote{A range of intermediate thresholds were also tested and verified to fit this overall trend but not shown here for clarity.}. Once again, PEF analyses outperform NWP at the initial time, and improvements are sustained longer for lower reflectivity thresholds (3--6 hours), which are more likely to arise from slowly evolving stratiform features, whereas higher thresholds contain a higher preponderance of unpredictable convective-scale features and improvements are exhausted by 3-4 hours. \fig \ref{pfss}a shows notable forecast improvements across all lead times at the 5 dBZ forecast verification threshold despite it falling outside our chosen tuning metrics. Note also that pFSS improvements are sustained for longer than point-based categorical metrics, indicating that much of the apparent skill loss at longer lead times in Fig.~\ref{rmse} arises due to displacement penalties rather than a genuine loss of useful forecast information.

\begin{figure*}[ht]
    \centering
    \noindent\includegraphics[width=\textwidth]{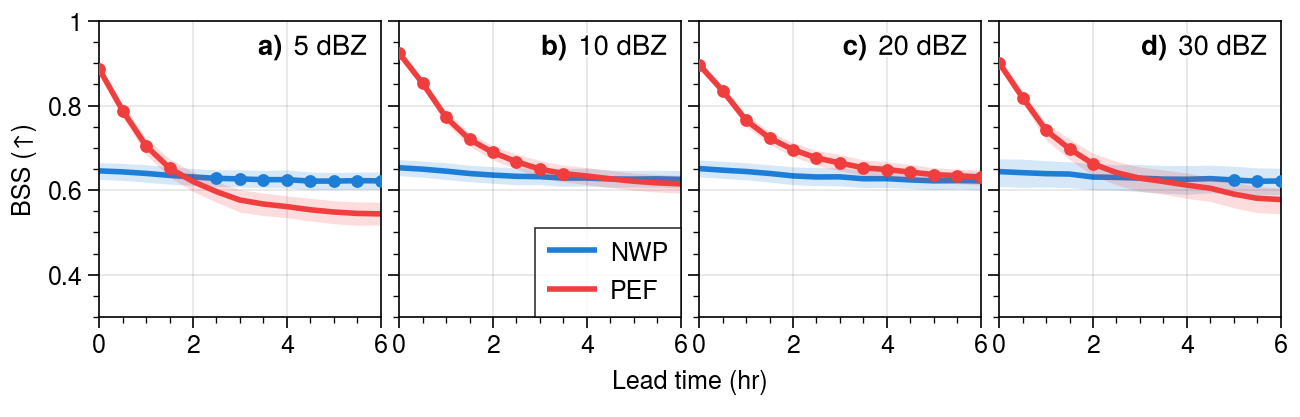}\\
    \caption{Neighborhood Brier skill score (BSS) versus lead time at $5$, $10$, $20$ and $30$ dBZ (25-cell window). Shaded bands indicate 99\% confidence intervals for NWP forecasts (blue), PEF analyses (red). Filled dots indicate lead times where one method performed significantly better than the other. Higher is better and values above zero indicate improvements over the no-skill climatology reference.}\label{brier}
\end{figure*}

\begin{figure*}[ht]
    \centering
    \noindent\includegraphics[width=\textwidth]{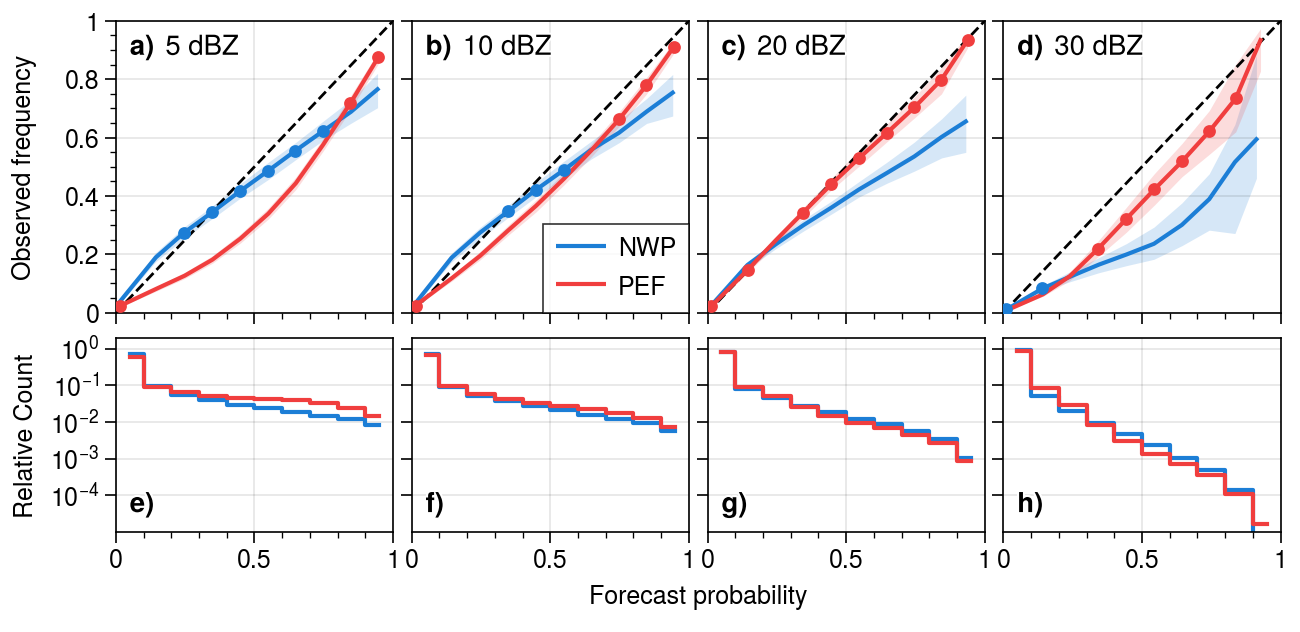}\\
    \caption{Reliability diagrams (top) and sharpness (bottom) at 5, 10, 20 and 30 dBZ (25-cell window), pooled over all forecast lead times. NWP forecast (blue), PEF analysis (red); the dashed diagonal represents perfect reliability and curves below it indicate overforecasting. Shaded bands indicate 99\% confidence intervals on the observed frequency. Filled dots indicate probability bins where one method is deemed significantly more reliable (closer to the 1:1 line) than the other. The bottom row is the sharpness diagram for each threshold: the relative frequency of forecasts in each probability bin (bin counts pooled over leads and normalized to sum to one), on a logarithmic axis.}\label{reliability}
\end{figure*}

The probabilistic accuracy and calibration of the forecasts are now summarized by the neighborhood Brier skill score (\fig \ref{brier}) and the reliability diagrams (\fig \ref{reliability}). Note that a range of window sizes were also calculated but not shown on \figs \ref{brier} \& \ref{reliability} because, like the pFSS in \fig \ref{pfss}, results for all window sizes showed the same qualitative behavior. The PEF begins with a high BSS of $\sim$0.9 for all thresholds and window sizes (25-cell window shown in \fig \ref{brier}), and reverts to the near-constant NWP skill between 2-6 hours depending on the threshold. Statistically significant improvements are observed for all thresholds at shorter lead times, while the PEF is shown to significantly outperform the NWP throughout the entire 6 hour forecast window in \fig \ref{brier}c. \figs \ref{reliability}b-d also show that this added forecast accuracy has also come with better ensemble calibration, although this is marginal for the 10 dBZ threshold in \fig \ref{reliability}b. These reliability plots show the PEF sits closer to the perfect 1:1 calibration line across most forecast probability bins for the 10, 20 and 30 dBZ thresholds. 

Importantly, both the 5 and 30 dBZ thresholds show modest, but statistically significant degradations of the Brier skill score at longer lead times, beyond roughly 2.5 and 5 hours, respectively (\figs \ref{brier}a,d; $99\%$ bootstrap confidence interval). The reliability diagrams (\figs \ref{reliability}a,d) reveal how these arise. At 5 dBZ the PEF analysis over-forecasts relative to the NWP, evidenced by its position below the 1:1 line and poorer performance relative to the NWP across the bulk of the probability range (\fig \ref{reliability}a). This most likely reflects the modulated analysis ensemble producing an excess of low reflectivities due to its broad localization smoothing length scale, and inflation settings. These panels and discussion were included for completeness, but poor calibration at 5 dBZ is tangential to our aim and tuning objective (\eq \eqref{J}) of producing a well-calibrated ensemble at the practically significant rainfall thresholds of 10, 20 and 30 dBZ. Applications with variables that require calibration across their full distribution should instead tune on additional Brier thresholds or an integrated CRPS.

A similar over-forecasting bias appears at the 30 dBZ threshold in the lowest forecast-probability bins ($0$--$0.25$), where the PEF is again significantly worse than the NWP (\fig \ref{reliability}d). The sharpness diagram (\fig \ref{reliability}h) shows that these bins contain the overwhelming majority of forecasts, so that even a small reliability bias there (and despite the significant wins at higher forecast probability bins), produces the long-lead 30 dBZ degradation \citep{Murphy73}. Unlike 5 dBZ, the 30 dBZ Brier score is part of our tuning objective, whose asymmetric cost (\eq \eqref{J}, $\beta=3$) penalizes such degradations threefold. The persistence of those significant long-lead time degradations despite the chosen metric implies that our tuning procedure was unable to find a single parameter combination yielding positive or neutral BSS improvements across all thresholds, neighborhood sizes and lead times simultaneously.

The pFSS, by contrast, shows no corresponding degradation at 5 or 30 dBZ. This is because the fractions skill score normalizes the neighborhood squared error by the forecast's own variance, and so rewards added forecast sharpness even when that sharpness is not matched by reliability \citep{Roberts08, Skok16}. In this way the fractions skill score can therefore be ``gamed'' toward higher values through sharpness alone, with no improvement in calibration. This is not possible for the Brier score, due to its status as a strictly proper scoring rule \citep{Gneiting07,Wilks11} minimized only when the forecast probabilities are reliable. For this reason, we prefer it for both tuning and verification.

\section{Conclusion and Future Directions}\label{conclusion}

We have shown that preemptive forecasting offers a computationally cheap means of improving short-range probabilistic guidance of rainfall (+0-6 hours). Forecast updates are calculated on the order of minutes (<5 min. in our implementation on a compute node with two 14-core Intel ‘Broadwell’ processors) rather than hours as in a full data-assimilation-forecast cycle of the physics-based, convection-permitting operational NWP system \citep{Rennie22}. This was shown by assimilating a radar reflectivity mosaic into a $24$-member lagged ACCESS-CE ensemble with flow-adaptive localization and a lead time-based observation-error inflation schedule. Important tunable parameters were set by minimizing a cost function that rewarded accuracy and calibration improvements relative to the background NWP forecast. Illustrative examples show that the resulting preemptive forecasting system can reposition and sharpen forecast precipitation features toward observations, and that these analysis improvements persist for many hours into the forecast window.

These claims were tested using nine weeks of five-hourly forecasts during periods of active precipitation. The preemptive forecasts were shown to deliver improvements relative to the NWP ensemble across almost all deterministic and probabilistic metrics considered. The gains are largest at the analysis time where the observations are already available, and decay with lead time until they are mostly indistinguishable from NWP by the end of the 6 hour forecast window. Displacement-tolerant neighborhood metrics retain skill longer than point-based scores, indicating there is some spatial uncertainty in the preemptive corrections. Neighborhood metrics in Table \ref{t:significance} show statistically significant improvements out to 3-6 hours for probabilistic fractions skill scores and 2-6 hours for Brier skill scores for various rainfall intensity thresholds. While it was beyond the scope of our methodologically-focused study to compare PEFs with existing radar-NWP blending approaches, a cursory comparison shows similar improvement horizons for statistical methods out to 3-6 hours \citep[\eg][]{Imhoff23}. 

\begin{table}[!ht]
\caption{Lead time out to which the pre-emptive forecast is statistically significantly better than the NWP baseline (99\% paired bootstrap confidence interval excluding zero), for the neighbourhood pFSS and Brier score at each reflectivity threshold ($w = 25$).}\label{t:significance}
\centering
\begin{tabular}{lll}
\toprule
Threshold & pFSS sig. improvement & Brier sig. improvement\\
\midrule
10 dBZ & to +6 h & to +3.5 h \\
20 dBZ & to +3 h & to +6 h \\
30 dBZ & to +4 h & to +2 h \\
\bottomrule
\end{tabular}
\end{table}

The principal limitation of using preemptive forecasts as a stand-alone nowcasting system concerns the underlying NWP ensemble. Preemptive forecasts are only provably optimal in cases where prior ensemble perturbations are small (\eg \eq \ref{tangent_linear}), and as a result, we expect that the relative improvements gained from PEFs should scale with the accuracy of the underlying ensemble forecast. In this case, the ACCESS-SYE ensemble showed uniform skill for radar reflectivity forecasts across lead times (0-6 hours), indicating very little convective-scale information is gained through its 4DVAR DA scheme. Its small, 12-member ensemble size (or 24 lagged members), and the large associated convective scale perturbations dictate that the value of a correction made at the analysis time is bounded to the few hours over which the prior covariances remain coherent. This is seen most clearly in rapidly evolving, nonlinear convection (\fig \ref{example2} and \fig \ref{brier}d), where preemptive corrections may actually degrade NWP forecasts. We have identified several avenues for future work that stem from these limitations and other modeling choices made in the current system:

\begin{itemize}
    \item \textbf{Bigger/better ensembles.} We aim to trial the technique with newer versions of the ACCESS model with more advanced convective-scale DA \citep{Rennie26}, and with larger systems offering more ensemble members. We expect both changes to lengthen the window over which preemptive forecast corrections offer improvements.
    \item \textbf{Non-Gaussian and bounded errors.} The clipped reflectivity used here is a non-negative, heavily skewed variable which we clip at a practical measurement sensitivity floor (see \fig \ref{prior_distro}). Clearly, the Gaussian-error assumptions underlying the ETKF are imperfect here, and we expect that future developments with bounded error models will also improve the analysis \citep[\eg][]{Bishop16}. 
    \item \textbf{Multiscale localization.} A single low-pass filter used in the localization procedure (cutoff $\ell\approx10$ grid cells) cannot perform the complex spectral transformations required to remove spurious correlations while simultaneously reproducing broad frontal bands and isolated cells. In our case this results in overforecasting small reflectivities (5 dBZ, \fig \ref{brier}a). Publication of recent work on performing scale-selective analysis updates using invertible filters to correct spectral content in the analysis will follow this manuscript. 
\end{itemize}

Having applied the preemptive forecasting idea to full-scale NWP model fields for the first time in this study, we also wish to comment on future practical usages. The method was introduced here in the context of correcting precipitation forecasts with radar data, and we have shown it may contain promise as a stand-alone post-processing technique for blending current observations with NWP forecasts for hydrological applications such as flood forecasting or for rapidly updating model guidance for severe weather warnings. This is attractive for physics-based models where running rapid update cycles remains computationally expensive, especially where existing convective-scale DA may be poor. Further, we also note that rapid remote sensing-based (radar or satellite) updates may also be applied to other model variables such as temperature and humidity; in this way, it may be possible to produce a time-varying analysis product for ingestion as an analysis increment in a standard DA window. This method may operate similarly to existing latent heat nudging approaches in which prognostic model variables are updated based on recent precipitation observations. Finally, we also note that the flow-adaptive localization procedure introduced here, when combined with an analysis ensemble size reduction procedure \citep[GETKF;][]{Bishop17}, has potential for use in a full-scale, cycling data assimilation scheme. 

\acks
This research was undertaken with the assistance of computational resources from the National Computational
Infrastructure (NCI Australia), an NCRIS-enabled capability supported by the Australian Government.

\section*{Funding information}
Open Access funding enabled and organized by CAUL and its Member Institutions. C. H. Bishop is partially supported by the Australian Research Council Centre of Excellence for 21st Century Weather (CE230100012).

\section*{Conflict of interest}
The authors declare no conflict of interest.

\section*{Permission to reproduce material from other sources}
All figures and tables in this manuscript were produced by the authors.

\section*{Author contributions}
\textbf{Jordan P. Brook:} conceptualization; methodology; software; formal analysis;
investigation; validation; visualization; writing -- original draft.
\textbf{Craig H. Bishop:} conceptualization; methodology; writing -- review and
editing.
\textbf{Alain Protat:} conceptualization; writing -- review and editing.

\section*{Data availability statement}
All ensemble forecast and radar data used in this study are publicly available at \url{https://thredds.nci.org.au/thredds/catalog/catalogs/wr45/catalog.html} and \url{https://thredds.nci.org.au/thredds/catalog/rq0/rainfields3/catalog.html}, respectively.

\bibliographystyle{plainnat}
\bibliography{references_clean}

\end{document}